\documentclass[amsmath,amssymb,showpacs,showkeywords,twocolumn]{revtex4}
\usepackage{graphicx}
\usepackage{times}
\usepackage{mathrsfs}
\usepackage{amsmath}
\usepackage{graphicx}
\usepackage{epsfig}
\usepackage{dcolumn}
\usepackage{bm}
\usepackage{multirow}
\usepackage{pdfpages}
\usepackage[utf8x]{inputenc}
\DeclareUnicodeCharacter{2212}{-}
\begin{document}
\title{Spin-resolved ballistic transport in three-terminal Zigzag Graphene Nanoribbon Device}
\author{Niharika Tamuli\footnote{tamuliniharika8@gmail.com} and Saumen Acharjee\footnote{saumenacharjee@dibru.ac.in} }
\affiliation{Department of Physics, Dibrugarh University, Dibrugarh 786 004, 
Assam, India}

\begin{abstract}
We investigate the spin-polarized ballistic transport in a three-terminal Zigzag graphene nanoribbon (ZGNR) device using a tight binding model, non-equilibrium Green function formalism within the Landauer–B\"{u}ttiker framework. We study the transmission spectrum, density of states, I–V characteristics, spin-resolved conductance and spin current by varying ribbon geometries and an out-of-plane Zeeman field.  In  absence of magnetization, transport is dominated by subband quantization and resonant edge states, with pronounced dependence on ribbon width and length while the introduction of a Zeeman field offers spin-selective transport and inducing half-metallic behavior, particularly in narrower ribbons, highlighting the interplay between quantum confinement, edge-localized states and spin-dependent interactions.  Moreover, we found Fabry–P\'{e}rot-like interference in conductance spectrum and bias-driven mode activation with strong spin filtering effects. The spin current is found to be tunable via magnetic field and gate voltage.  Also, it remains stable under thermal fluctuations, demonstrating suitability for room-temperature operation. Finally, the energy and width dependence of the Fano factor reveals distinct quantum interference features and spin-polarized transport signatures. These findings indicate the potential of the three-terminal ZGNR based device for scalable and gate-controllable spintronic applications.
\end{abstract}

\pacs{71.70.Ej, 72.80.Vp, 73.23.−b, 81.05.ue, 85.30.Tv}
\maketitle

\section{Introduction}
Spintronics, which harnesses the spin degree of freedom alongside charge, has emerged as a promising paradigm in nanoelectronics to enable next-generation logic and memory applications \cite{zutic2004,hirohata2020,han2018,ding2020,datta2018}.  The central challenge in this domain lies in the realization of spin-polarized transport with minimal dissipation while maintaining long coherence length \cite{datta2018}. This limitation can be easily overcome in ballistic regime, where charge and spin carriers can travel freely without scattering \cite{vila2020,guimaraes2012,rao2023,hartel2023,roy2023,borge2017}.  Ballistic spin transport not only minimizes energy loss but also preserves quantum phase information.  Thus, it enables enhanced control over spin dynamics and coherent manipulation \cite{kavokin2013,mishra2021,acharjee2024}. In this context, low-dimensional materials that support spin-polarized currents are of particular interest. Among these, graphene and its derivatives have received significant attention because of their exceptionally high carrier mobility, long spin relaxation time and atomically thin geometry \cite{kunstmann2011,lee2009}. However, monolayer pristine graphene is a gapless semimetal and has spin-degenerate bands \cite{spyrou2014,novoselov2007}. Thus, its effectiveness in field-effect transistors and spin-filtering applications is limited. To overcome these
limitations, quasi-one-dimensional variants such as graphene
nanoribbons (GNRs) have given significant attention \cite{zhang2010,kumar2023,saroka2014}.  GNRs are nanometer-wide graphene strips that exhibit quantum confinement and edge-dependent properties.  In addition, they have the ability to form bandgapsand support spin-polarized states \cite{tombros2007,dutta2010,pratap2017,tian2023,liu2010}. GNRs are classified into two subclasses based on their edge geometries: Armchair edged graphene nanoribbons (AGNR) and Zigzag edged graphene nanoribbons (ZGNR) \cite{zhang2010,kumar2023,saroka2014}.

ZGNRs are of particular interest because of their intriguing edge terminations. Unlike AGNRs, with a size-dependent bandgap, ZGNRs have localized edge states that cause almost flat bands at the Fermi energy \cite{kunstmann2011,  lee2009}. In addition, owing to the strong Coulomb interactions, the edge states spontaneously spin-polarize, resulting in ferromagnetic coupling on both edges while antiferromagnetic coupling between the two opposite edges \cite{kan2012, zhang2021}. Importantly, this intrinsic edge magnetism has the potential to generate spin-polarized currents even in the absence of an external magnetic field, thereby opening the way for designing spintronic devices made entirely of graphene. It is found that a ZGNR can act as a spin filter or spin valve in two-terminal devices, depending on the edge magnetization orientation and bias polarity \cite{ganguly2018n}.  However, two-terminal configurations are constrained in their ability to explore more intricate phenomena such as nonlocal voltage generation, asymmetric transmission channels or spin-dependent channel resolved current \cite{dankert2014,farghadan2015,tralle2005}. This limitation motivates the exploration of multiterminal setups, which introduce an additional degree of control over spin and charge flow and thereby allow for richer physical effects \cite{farghadan2015}. Three-terminal mesoscopic devices provide functionalities beyond those of common two-terminal configurations, such as current rectification, nonlocal resistance, and spin-charge separation. In spintronic systems based on ZGNR, such geometries utilize the spin-polarized edge states to access nonreciprocal transport, spin filtering and gate-controllable spin manipulation \cite{tralle2005,pastawski1991, kuvcera1991}. The asymmetric coupling between the nanoribbon and multiple leads induces spatial and spin asymmetries in the transmission spectrum and provide control of spin selectivity through geometric or electrostatic tuning.

Although substantial theoretical and experimental efforts have been made to understand spin-polarized transport in GNRs, most studies have focused on two-terminal configurations, primarily considering the antiferromagnetic ordering of localized edge states in ZGNR \cite{son2006, dutta2008, jung2009, ezawa2006,soriano2010}. These systems utilize intrinsic spin polarization at the edges to realize GNR based spin valves \cite{wang2007} and spin filters \cite{hod2007,hod2008} based on quasi-flat bands at the Fermi level originating from edge-localized states \cite{wakabayashi2010}. In multiterminal geometries, one can manipulate individual terminal voltages independently to realize spin switching, nonreciprocal conductance effects, and directional spin flow. Three-terminal graphene devices with Rashba spin-orbit coupling have been found to create gate-controllable spin currents and display simple spin logic \cite{ganguly2018, manchon2015}. Moreover, spin filtering by Coulomb interaction has also been realized in multiterminal devices with quantum dots and extended nanoribbon segments \cite{cresti2014}. Furthermore, recent studies demonstrate the potential of three-terminal ZGNR junctions for spintronic applications, enabling valley-controlled dual-channel charge pumping and the conversion of pure spin currents into measurable charge currents via valley-valve effects \cite{zhang2013,zhang2017}.  However, these proposals generally rely on extrinsic spin selectivity mechanisms such as proximity-induced spin–orbit fields or gate-defined quantum confinement instead of intrinsic ones. Studies focusing on ballistic, interaction-driven spin transport in pristine three-terminal ZGNRs are still limited. Nevertheless, there is still insufficient theoretical analysis on ballistic spin transport in three-terminal ZGNRs with quantum coherence, geometry-induced interference and edge-selective channeling. In this work, we address this gap by employing a tight-binding model and the non-equilibrium Green function (NEGF) formalism to compute spin-resolved transmission in a three-terminal ZGNR device. We demonstrate ballistic transport and tunable spin polarization controlled by ZGNR geometry, magnetization and bias direction. Our results advance the understanding of coherent spin manipulation in graphene nanostructures and support the design of spin-functional quantum logic circuits.

The remainder of the paper is organized as follows. In Section II, we present our model and theoretical formalism based on a tight binding Hamiltonian for a three-terminal ZGNR device. Spin polarized transport properties have been studied using the NEGF formalism along with Landauer-B\"{u}ttiker formula. In Section III, we present and analyze the numerical results, highlighting key features of spin-dependent transport in three terminal ZGNR based devices. Finally, Section IV summarizes our key findings and outline the potential avenues for experimental realization and future developments in ZGNR based devices.

\section{Model and Formulation} 
\begin{figure}[t]
\centering
\centerline{
\includegraphics[scale=0.37]{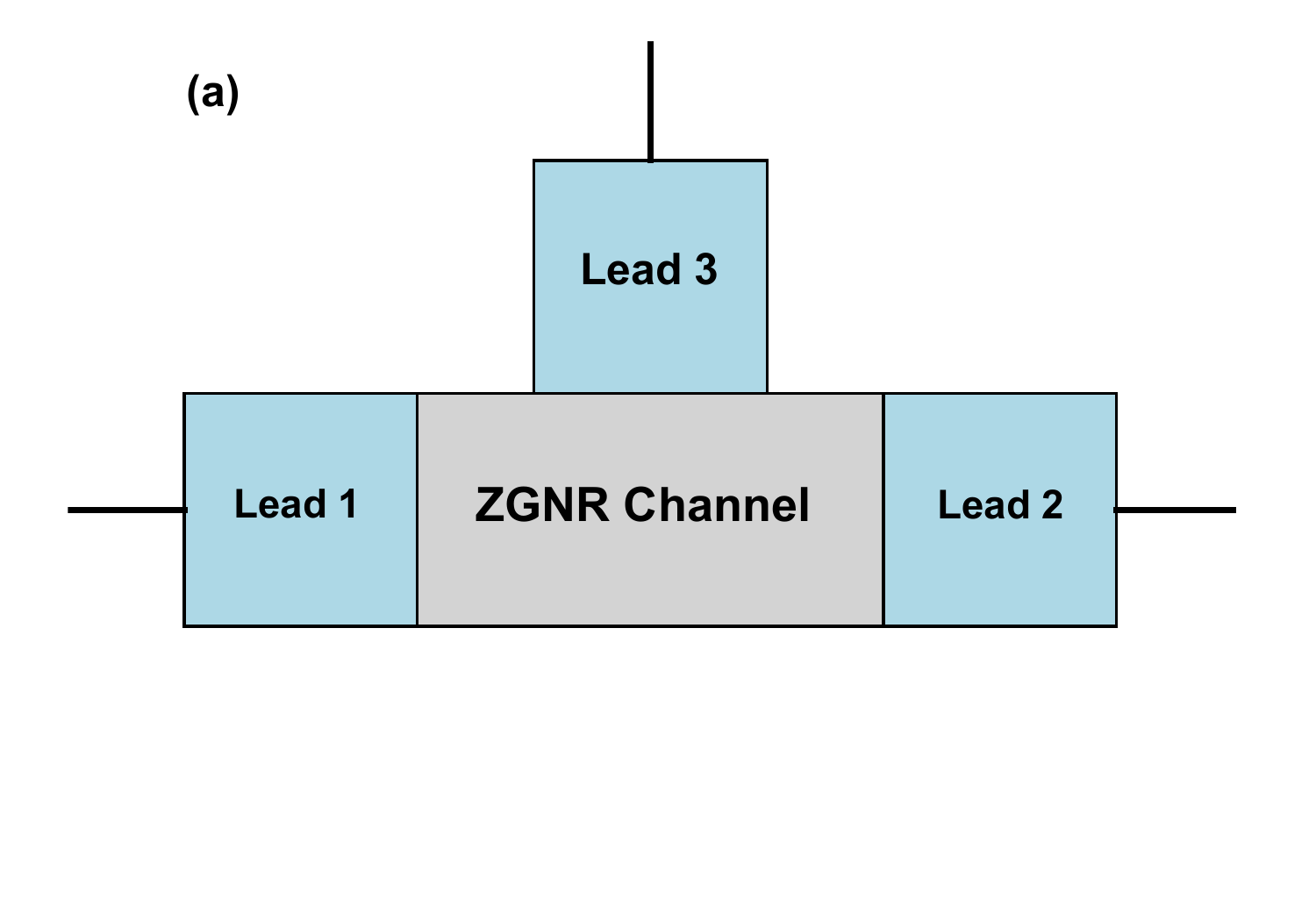}}
\vspace{-15mm}
\centerline{ 
\includegraphics[scale=0.27]{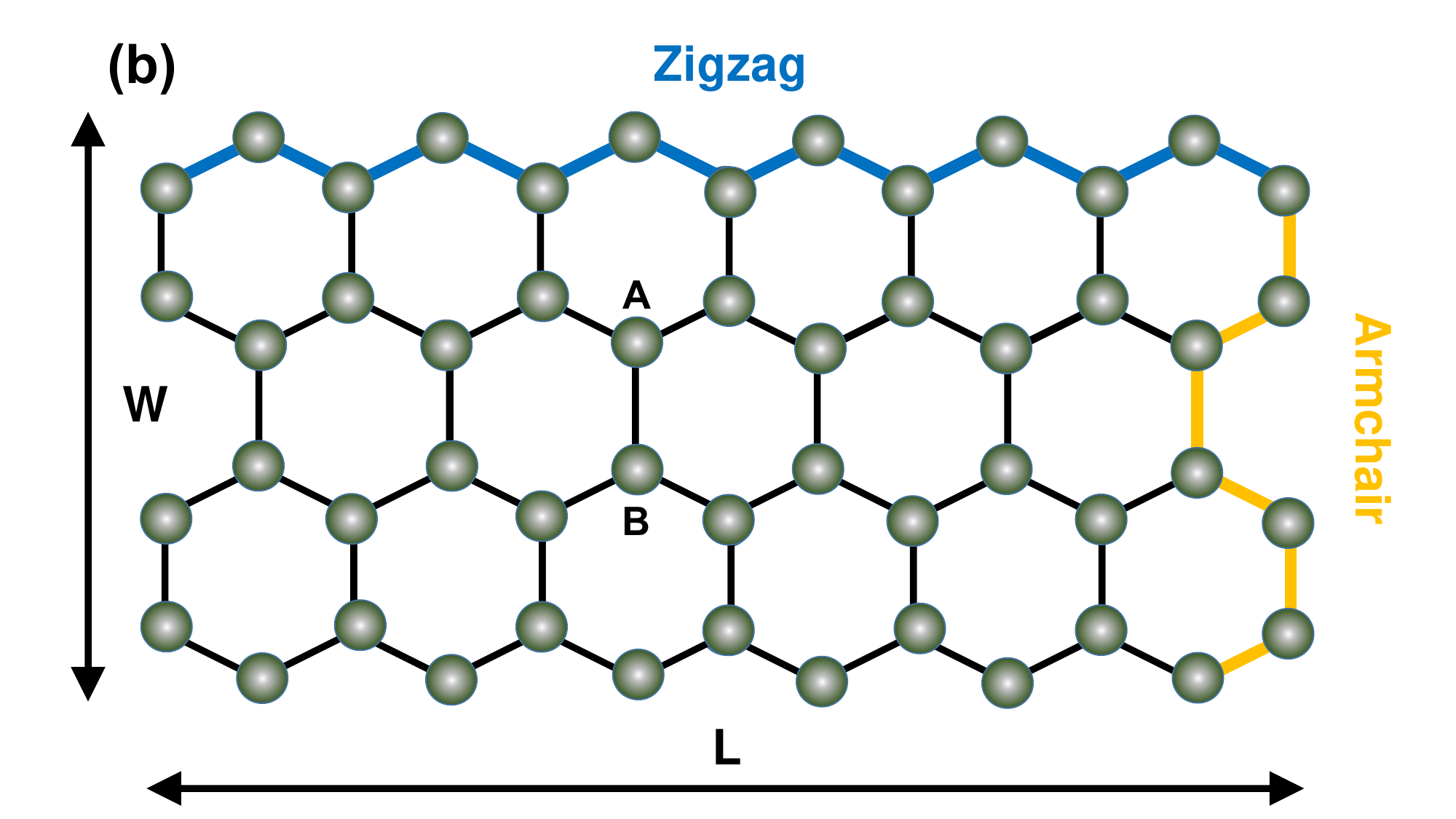}}
\centerline{ 
\includegraphics[scale=0.36]{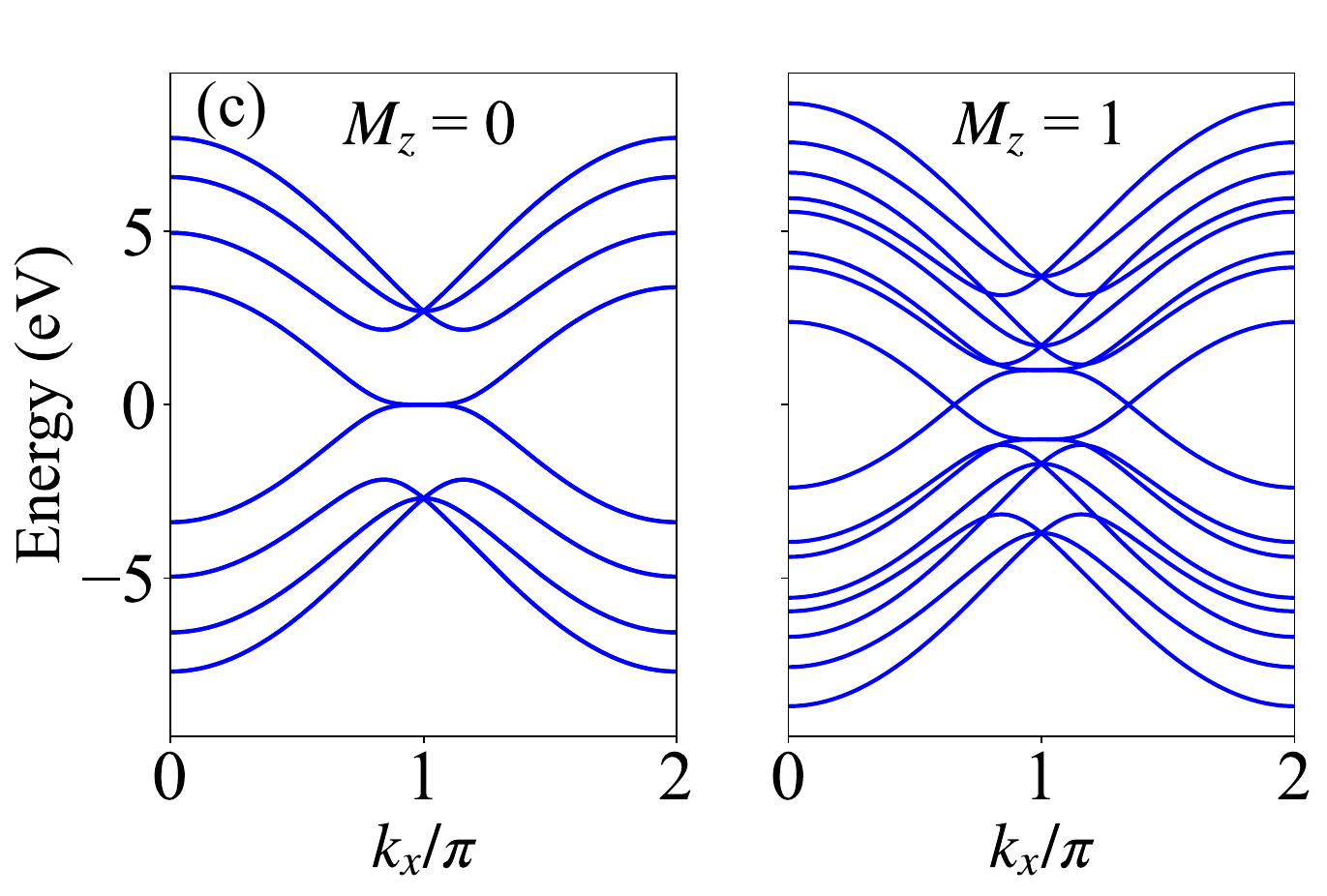}}
\caption{(Top Panel) Block diagram of a three-terminal ZGNR device. ZGNR is considered as a channel is connected to three metallic leads:
Lead 1 (source) at the left edge, Lead 2 (drain) at the right
edge and Lead 3, which is attached perpendicularly to the
top center of the channel, functioning as a voltage probe. (Middle Panel) A cross-section of ZGNR. (Bottom Panel) Energy band structure for ZGNR for $M_z$ = 0 (left) and 1 (right).}
\label{fig1}
\end{figure}

We consider a three terminal ZGNR based device as shown in Fig.\ref{fig1}. The central region consists of a finite-width ZGNR, which serves as the active transport channel.This channel is connected to three metallic leads: Lead 1 (source) on the left edge, Lead 2 (drain) on the right edge and Lead 3, which is attached perpendicularly to the top center of the channel and can function as a voltage probe. In graphene, each carbon atom forms three $\sigma$-bonds with neighboring carbon atoms on the two-dimensional plane while the fourth electron occupies a $p_z$ orbital oriented perpendicular to the plane, forming delocalized $\pi$-bond above and below the graphene sheet. These highly mobile $\pi$ electrons are responsible for electronic conduction. The $\pi$ orbitals overlap and enhance the carbon-carbon (C–C) bonding in graphene. Fundamentally, the electronic properties of graphene are dictated by the bonding ($\pi$) and anti-bonding ($\pi^\ast$) orbitals, which form the valence and conduction bands \cite{castro2009}.

To begin with, we consider a spinless tight-binding model on a honeycomb lattice, where each lattice hosts a single $\pi$ orbital and electrons can hop to their nearest neighbors. The effective tight-binding Hamiltonianfor ZGNR can be written as \cite{nakada1996,hancock2010} 
\begin{equation}
\hat{\mathcal{H}}_0 = -t \sum_{\langle i,j \rangle} c_i^\dagger c_j + \text{H.c.}
\label{eq1}
\end{equation}
where, $c_i^\dagger$ ($c_j$) creates (annihilates) an electron at site $i$ ($j$) and the summation $\langle i,j \rangle$ runs over nearest-neighbor pairs with the hopping amplitude $t = 2.7$ eV. 

Considering the spin degree of freedom, the total Hamiltonian of the system can be written as \cite{datta1997}
\begin{equation}
\hat{\mathcal{H}} = \hat{\mathcal{H}}_{\text{hop}} + \hat{\mathcal{H}}_{\text{Zeeman}} + \hat{\mathcal{H}}_{\text{gate}}+\hat{\mathcal{H}}^{(\alpha)}_{\text{lead}}+\hat{\mathcal{H}}^{(\alpha)}_{\text{LD}}
\label{eq2}
\end{equation}
where, the first term corresponds to the hopping between nearest-neighbor sites and is given by
\begin{equation}
\hat{\mathcal{H}}_{\text{hop}} = -t \sum_{\langle i,j \rangle} \sum_{\sigma} \left( c_{i\sigma}^\dagger c_{j\sigma} + \text{H.c.} \right)
\label{eq3}
\end{equation}
where, $c_{i\sigma}^\dagger$ creates an electron with spin $\sigma$ at the lattice site $i$.

The Zeeman interaction due to an external magnetic field along the $z$-direction can be expressed as \cite{datta2015}
\begin{equation}
\hat{\mathcal{H}}_{\text{Zeeman}} = M_z \sum_i \left( c_{i\uparrow}^\dagger c_{i\uparrow} - c_{i\downarrow}^\dagger c_{i\downarrow} \right)
\label{eq4}
\end{equation}
where $M_z$ denotes the magnitude of the Zeeman energy measured in terms of Fermi energy.

The third term represents the effect of a local electrostatic potential or gate voltage and can be given by
\begin{equation}
\hat{\mathcal{H}}_{\text{gate}} = \sum_i V_i \sum_{\sigma} c_{i\sigma}^\dagger c_{i\sigma}
\label{eq5}
\end{equation}
where $V_i$ represents the on-site gate potential at site $i$. Diagonalizing the Hamiltonian in Eq. (\ref{eq2}), we can obtain the energy band structure for ZGNR. Fig. \ref{fig1}(c), shows the electronic band structure of ZGNR for $M_z = 0$ (left) and $M_z = 1$ (right). In absence of Zeeman field, the spin-degenerate bands show nearly flat edge states near the Fermi level due to localized edge modes which is the typical characteristic of ZGNRs. In the presence of a Zeeman field ($M_z = 1$), the spin degeneracy is lifted, leading to spin-split bands. This spin polarization originates from the Zeeman interaction, which energetically shifts the spin-up and spin-down states in opposite directions, enabling spin-dependent transport in the three-terminal setup.

The Hamiltonian for each semi-infinite lead can be modeled as
\begin{equation}
\hat{\mathcal{H}}_{\text{lead}}^{(\alpha)} = -t \sum_{\langle i,j \rangle \in \alpha} \sum_{\sigma}
\left( c_{i\sigma}^{\dagger} c_{j\sigma} + \text{H.c.} \right)
\label{eq6}
\end{equation}
where, $\alpha \in \{\text{L, R, T}\}$ represents the left, right and top leads of the proposed system. The last term of Eq. (\ref{eq2}) corresponds to the interaction of the leads with the device and can be written as

\begin{equation}
\hat{\mathcal{H}}_{\text{LD}}^{(\alpha)} = -t_c \sum_{\langle i,j \rangle} \sum_{\sigma}
\left( c_{i\sigma}^{\dagger(\alpha)} c_{j\sigma} + \text{H.c.} \right)
\label{eq7}
\end{equation}
where, $t_c$ is the coupling strength between the lead $\alpha$ and the device region, and $\langle i,j \rangle$ indicates hopping between the lead boundary sites and the central region.

\subsection{Density of States}
The transmission coefficients and density of states (DOS) are computed by evaluating the retarded Green function of the central region. The retarded Green function $\mathcal{G}^r(E)$ of the central region is defined as \cite{datta1997,CCaroli_1971,datta2015}
\begin{equation}
\mathcal{G}^r(E) = \left[ E \hat{I} - \hat{\mathcal{H}} - \sum_{\alpha} \Sigma_{\alpha}^r(E) \right]^{-1}
\label{eq8}
\end{equation}
where, $E$ is the total energy of the system, $\hat{I}$ is the identity matrix and $\Sigma_{\alpha}^r(E)$ are the retarded self-energy due to lead $\alpha$. The self-energy $\Sigma_\alpha^r(E)$ is an energy-dependent term that encapsulates the effect of lead $\alpha$ on the central region. It modifies both the real and imaginary parts of the Hamiltonian and leads to level broadening and energy shifts. The broadening function $\Gamma_{\alpha}(E)$ due to the coupling with lead $\alpha$ is  given by \cite{datta1997}
\begin{equation}
\Gamma_{\alpha}(E) = i \left[ \Sigma_{\alpha}^r(E) - \Sigma_{\alpha}^a(E) \right]
\label{eq9}
\end{equation}
where, the advanced self-energy follows the condition $
\Sigma_{\alpha}^a(E) = \left\{ \Sigma_{\alpha}^r(E) \right\}^\dagger$. The transmission probability from lead $\beta$ to lead $\alpha$ at energy $E$ is given by \cite{datta1997,CCaroli_1971} 
\begin{equation}
T_{\alpha \beta}(E) = \mathrm{Tr} \left[ \Gamma_{\alpha}(E) \, \mathcal{G}^r(E) \, \Gamma_{\beta}(E) \, \mathcal{G}^a(E) \right]
\label{eq10}
\end{equation}
where the advanced Green function follows the condition
$\mathcal{G}^a(E) = \left\{\mathcal{G}^r(E) \right\}^\dagger$
Thus, the total DOS of the system can be given by 
\begin{equation}
\rho(E) = -\frac{1}{\pi} \, \mathrm{Im} \left[ \mathrm{Tr} \, \mathcal{G}^r(E) \right]
\label{eq11}
\end{equation}
For a spatially resolved system, the retarded Green function can be defined as 
\begin{equation}
\mathcal{G}^r(\mathbf{r}, \mathbf{r}'; E) = \langle \mathbf{r} | \left( E + i\eta - \hat{\mathcal{H}} \right)^{-1} | \mathbf{r}' \rangle,
\label{eq12}
\end{equation}
where, $\eta \to 0^+$ is a positive infinitesimal that ensure the causality of the retarded Green function. The states $|\mathbf{r}\rangle$ and $|\mathbf{r}'\rangle$ are position eigenstates at spatial coordinates $\mathbf{r}$ and $\mathbf{r}'$ respectively. 
So, the local density of states (LDOS) at position $\mathbf{r}$ can be written as
\begin{equation}
\rho(\mathbf{r}, E) = -\frac{1}{\pi} \, \mathrm{Im} \left[ \mathcal{G}^r(\mathbf{r}, \mathbf{r}'; E) \right]
\label{eq13}
\end{equation}

\subsection{Transport formalism}
The current flowing from lead $\alpha$ to lead $\beta$ can be obtained by using the Landauer–B\"{u}ttiker formula \cite{cornean2005,landauer1957,nemnes2004}:
\begin{equation}
I_\alpha = \frac{e}{h} \int_{-\infty}^{\infty} T_{\alpha\beta}(E) \left[ f_\alpha(E) - f_\beta(E) \right] dE
\label{eq14}
\end{equation}
where, $e$ is the elementary charge, $h$ is Planck’s constant and $f_\alpha(E) = \left[1 + \exp\left( \frac{E - \mu_\alpha}{k_B T} \right)\right]^{-1}$ is the Fermi–Dirac distribution function for lead $\alpha$ with chemical potential $\mu_\alpha$.

The differential conductance at small bias can be obtained by using the relation \cite{datta1997,CCaroli_1971,datta2015}
\begin{equation}
G_{\alpha\beta} = \frac{e^2}{h} \int_{-\infty}^{\infty} \left( - \frac{\partial f(E)}{\partial E} \right) T_{\alpha\beta}(E) \, dE
\label{eq15}
\end{equation}
Taking into account the spin degree of freedom, the spin-resolved current flowing into lead $\alpha$ with spin $\sigma$ is given by
\begin{equation}
I_\alpha^{\sigma} = \frac{e}{h} \sum_{\beta \neq \alpha} \int_{-\infty}^{\infty} \left[ T_{\beta\alpha}^{\sigma}(E) f_\alpha(E) - T_{\alpha\beta}^{\sigma}(E) f_\beta(E) \right] \, dE
\label{eq16}
\end{equation}
where, $f_\alpha(E)$ and $f_\beta(E)$ are the Fermi-Dirac distribution functions in leads $\alpha$ and $\beta$ respectively. Thus, the total charge current ($I_C$) and the spin current ($I_S$) can be obtained using the relations
\begin{align}
I_C &= I_\alpha^{\uparrow} + I_\alpha^{\downarrow}
\label{eq17}\\
I_S &= I_\alpha^{\uparrow} - I_\alpha^{\downarrow}
\label{eq18}
\end{align}
Eqs. (\ref{eq17}) and (\ref{eq18}) provide the fundamental basis for computing both charge and spin transport in the multi-terminal divice. Using these relations along with the energy-resolved transmission functions, we evaluate the spin-resolved and net charge currents under various bias and temperature conditions. 

\begin{figure}[t]
\centerline{
\includegraphics[scale=0.28]{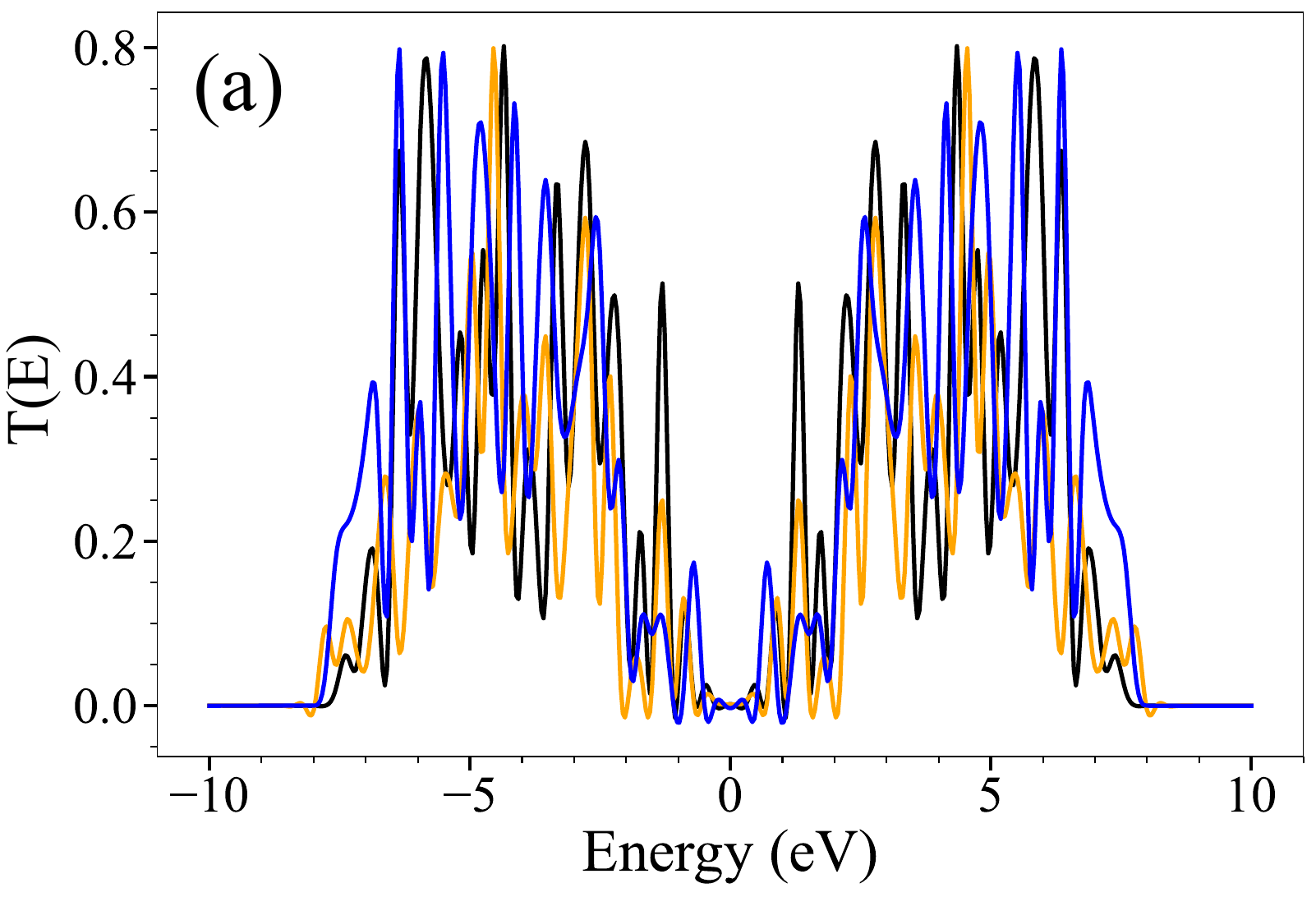}}
\hspace{0.1mm}
\centerline{
\includegraphics[scale=0.28]{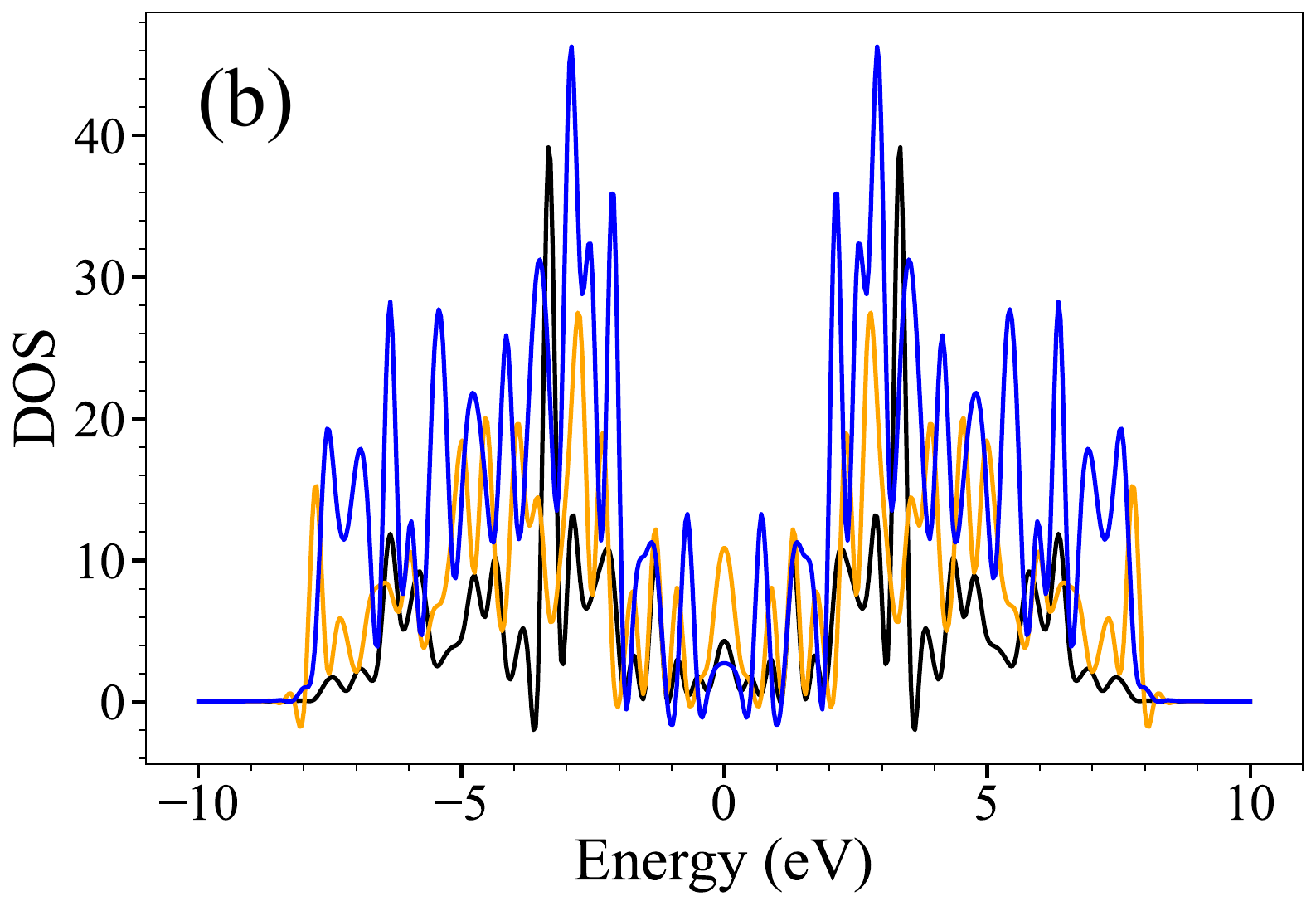}}
\centerline{
\includegraphics[scale=0.28]{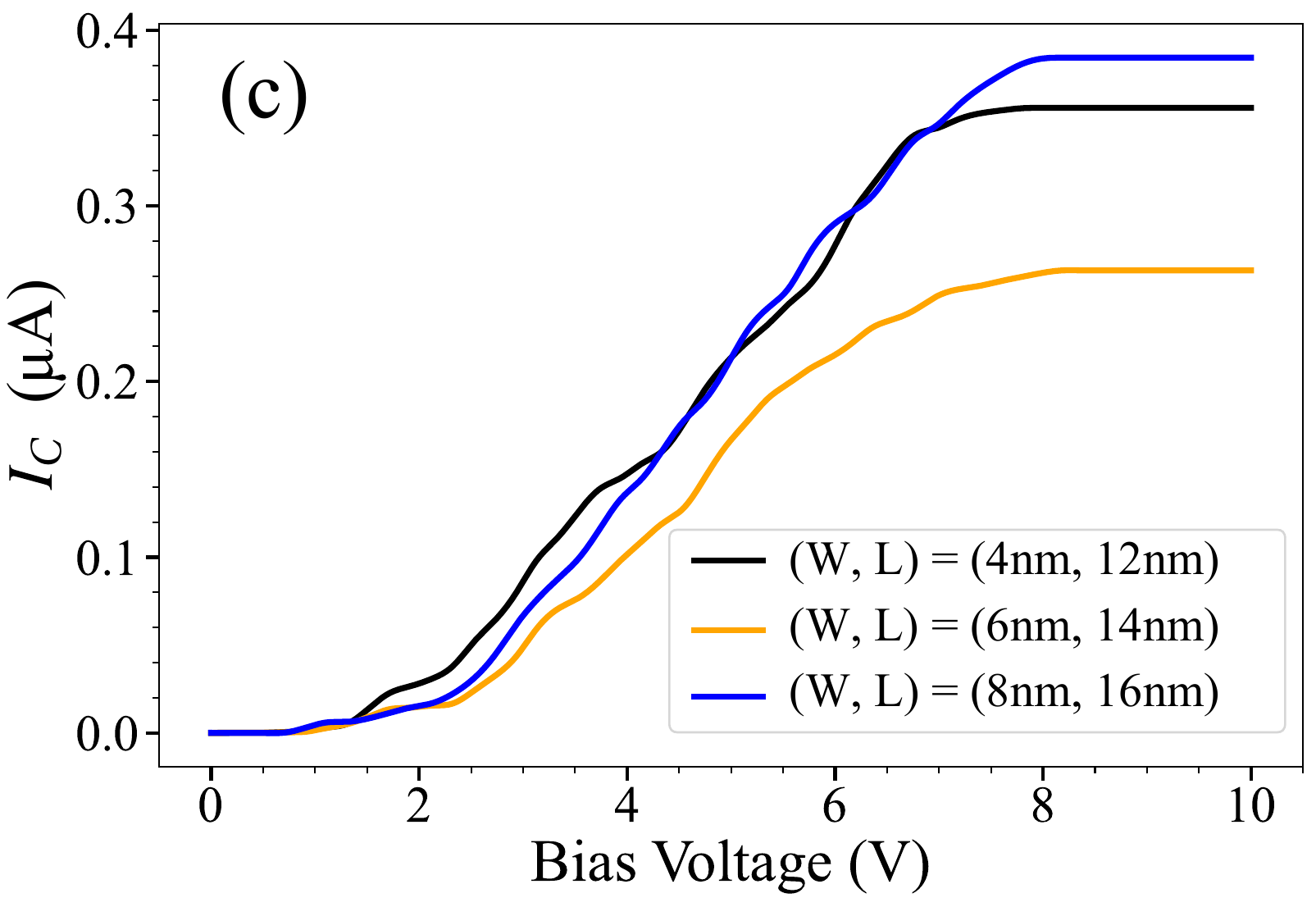}}
\caption{(a) Transmission coefficient and (b) Density of states (DOS) as functions of energy for ZGNRs with varying widths and lengths. (c) I-V characteristics of the three terminal ZGNR device for different width and length.}
\label{fig2}
\end{figure}

\section{Results and Discussion}
\subsection{Transport properties in absence of Magnetization}
We present and analyze our numerical results of the transmission coefficient, DOS and I-V characteristics in
Fig. \ref{fig2} for different ZGNR dimensions in the absence of magnetization. Fig. \ref{fig2}(a) illustrates the variation of the energy-resolved transmission coefficient $T(E)$ for the ZGNR device for three different ribbon geometries, characterized by their width ($W$) and length ($L$). We found that there exist prominent transmission peaks in the low-energy region around the Fermi level associated with edge-localized states \cite{kunstmann2011}. The transmission spectrum becomes denser and more structured as the width increases. This is due to the increase in the number of transverse modes available in the channel that contribute to conduction \cite{su2018}. In a similar way, an increase in the device length introduces more scattering centers, which may slightly suppress or modulate transmission, but the dominant effect arises from the increased width. The quantized peaks signify the resonant transport through available conduction channels and the oscillations originate from quantum interference due to coherent transport in the ballistic regime. The corresponding DOS are presented in Fig. \ref{fig2}(b) for the same ZGNR geometries. The DOS profile shows prominent and symmetric van Hove singularities (VHS) typical of quasi-one-dimensional systems with more pronounced features in the wider ribbons. The flat bands near the Fermi energy are associated with localized edge states for the ZGNR, leading to sharp peaks in the DOS in the low-energy region. The increase in ribbon width allow more subbands because of the increase in the number of carbon atoms in the transverse direction. This results in a more complex DOS profile with significantly increased numbers of VHS. These DOS characteristics are directly correlated with the available conducting channels and the transmission behavior in panel Fig. \ref{fig2}(a).

\begin{figure*}[t]
\centerline{ 
\includegraphics[scale=0.28]{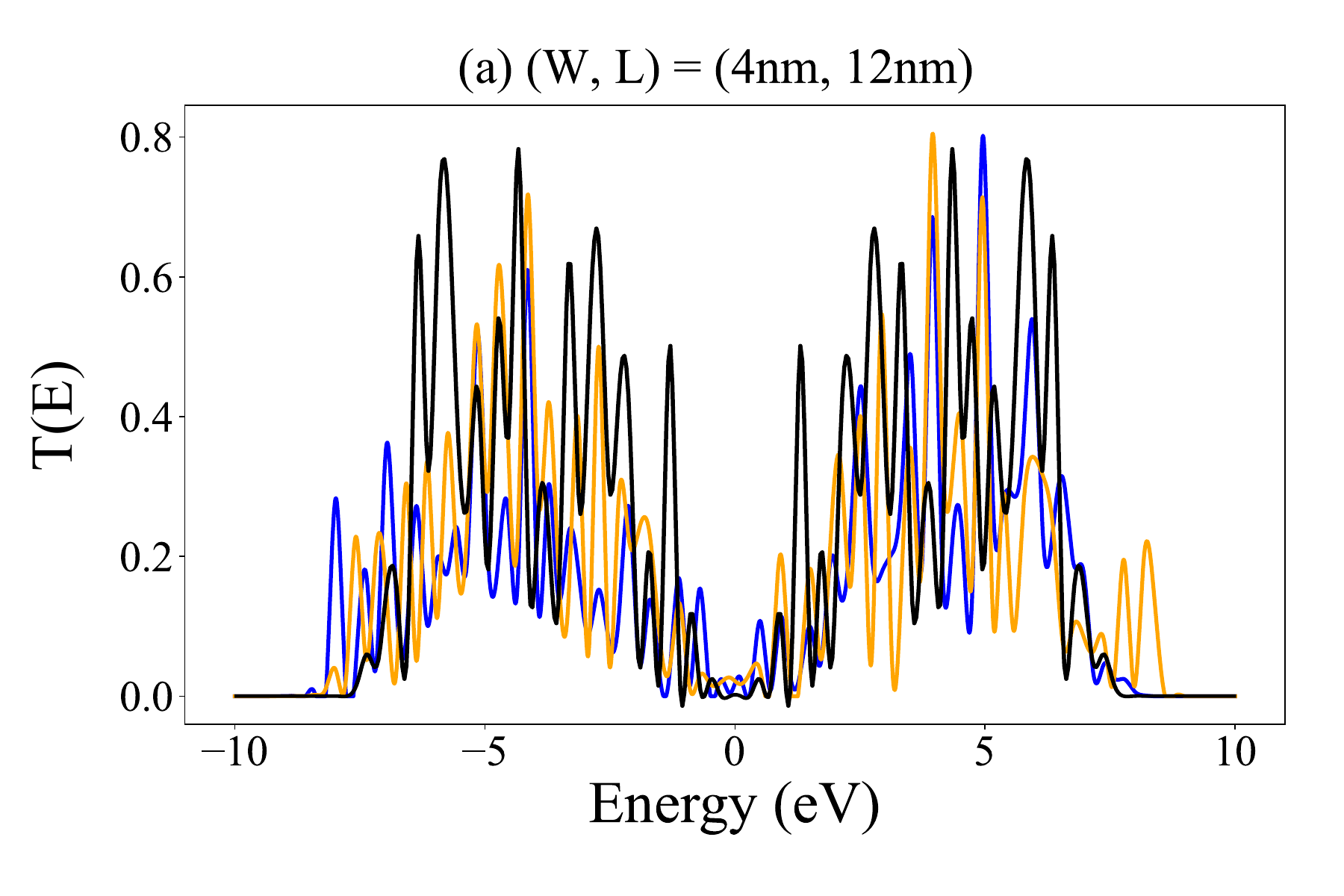}
\hspace{-5.mm}
\includegraphics[scale=0.28]{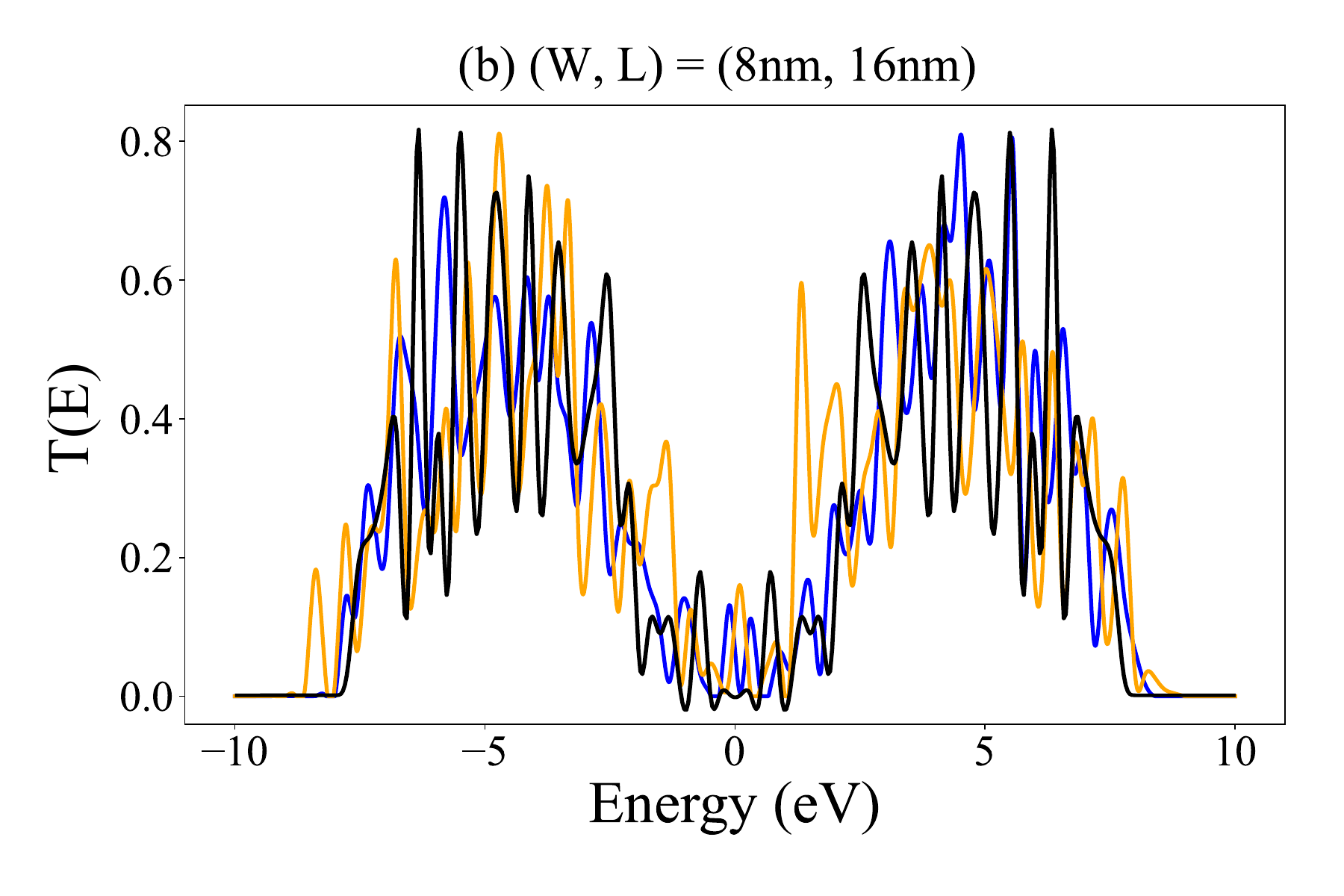}
}
\vspace{-5.mm}
\centerline{ 
\includegraphics[scale=0.28]{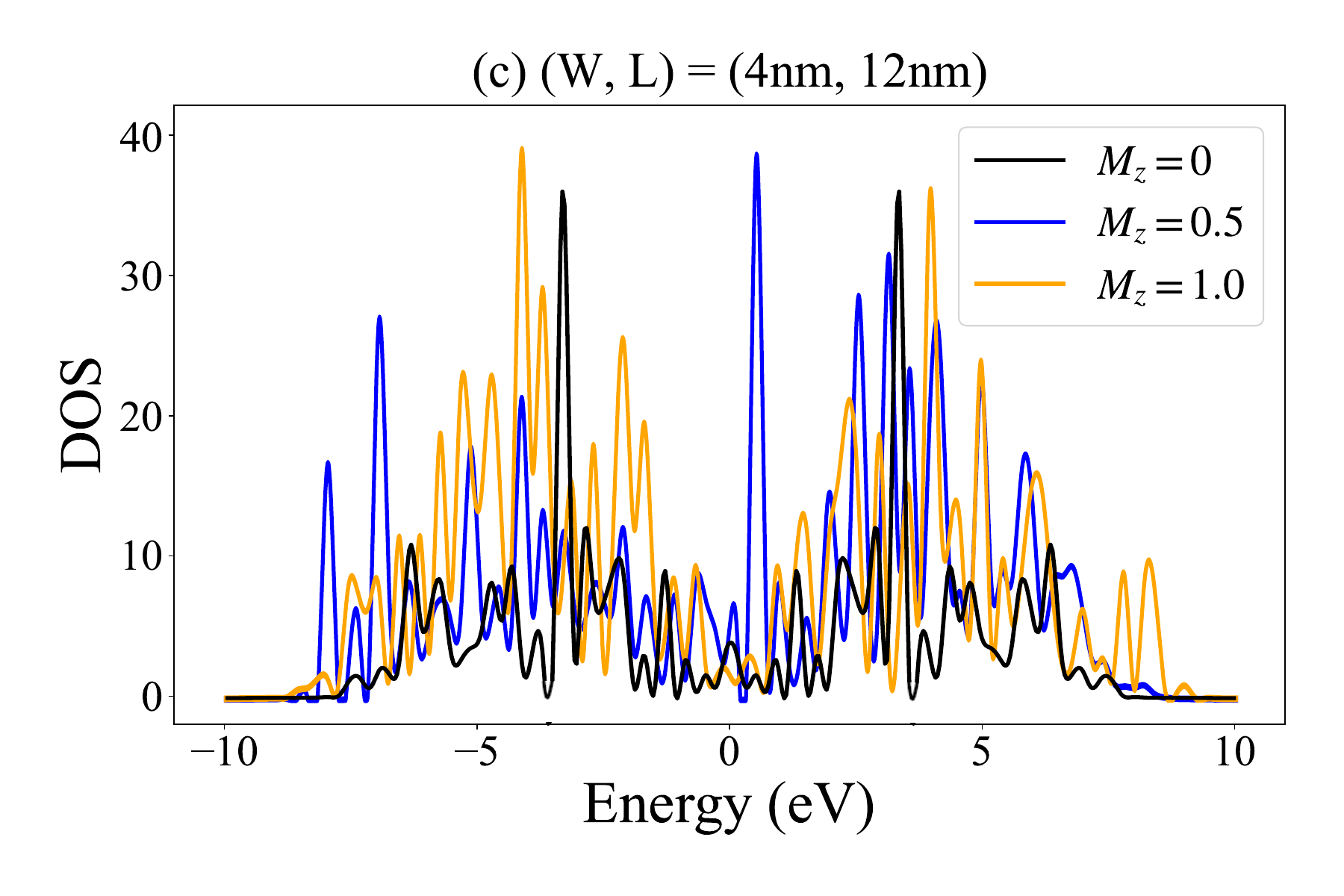}
\hspace{-5.mm}
\includegraphics[scale=0.28]{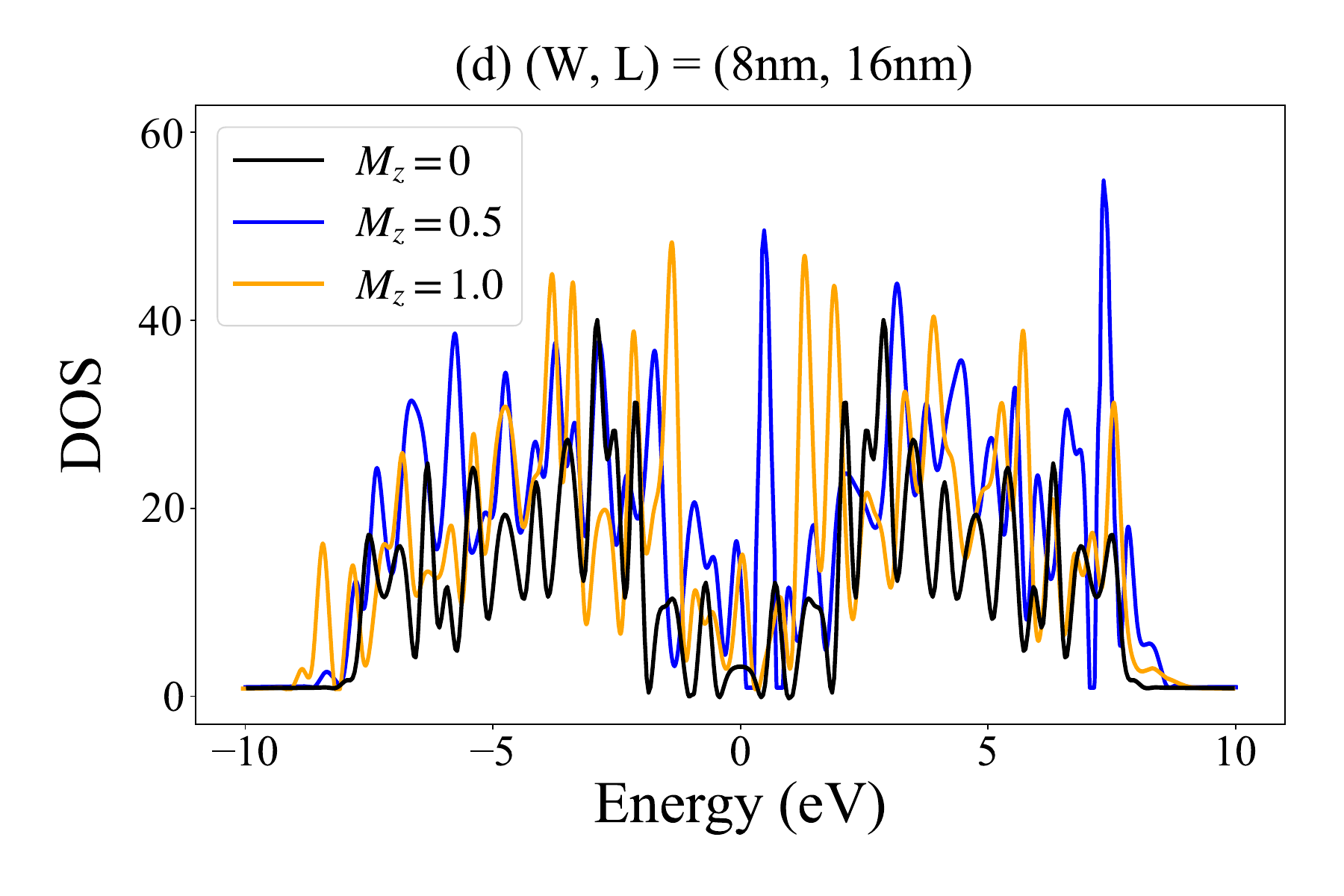}}
\caption{{(Top panel) Transmission coefficient as a function of energy for different magnetization strengths  considering  (a) $(W, L) = (4 \mathrm{nm}, 12 \mathrm{nm})$ and (b) $(W, L) = (8 \mathrm{nm}, 16 \mathrm{nm})$ of ZGNR. (Bottom panel) Variation of the DOS with energy for different magnetization strengths  considering  (c) $(W, L) = (4 \mathrm{nm}, 12 \mathrm{nm})$ and (d) $(W, L) = (8 \mathrm{nm}, 16 \mathrm{nm})$ of ZGNR.  }}
\label{fig3}
\end{figure*}
Fig. \ref{fig2}(c) shows the I–V characteristics of the ZGNR system in absence of magnetization within a bias window determined by the applied voltage V. We found that the current remains negligibly small in low bias voltage region ($V < 2$V). This is because of the presence of a transport gap near the Fermi level, which is attributed to the suppression of transmission by edge-state interference and quantum confinement effects. With an increase in the bias voltage ($V > 2$V), the current grows non-linearly, indicating the activation of additional conduction channels as energy levels fall within the bias window. It is observed that wider and longer ribbons show enhance current response due to to a higher density of available propagating modes and an increased transmission probability. In particular, the $(W, L) = (8 \mathrm{nm}, 16 \mathrm{nm})$ configuration exhibits the steepest slope, reflecting its superior conductive capacity due to the presence of multiple transverse subbands. The saturation observed beyond $V \sim 8$ V suggests the finite extent of the conduction window in the given energy range. Thus, the I-V characteristics highlight the interplay between ribbon geometry, mode quantization and coherent transport in a nanoscale ZGNR device.

\subsection{Transport properties in presence of Magnetization}
The interplay between magnetism and quantum transport in low-dimensional systems has been a matter of great interest due to its pivotal role in the advancement of spintronics and quantum information technologies. In this context, GNRs provide a controllable platform for investigating spin-resolved transport phenomena because of their outstanding electronic properties and controllable edge states. A particularly intriguing modification of electronic properties of GNRs arises with the introduction of a Zeeman field, which serves as an effective out-of-plane magnetic exchange field that removes the spin degeneracy of electronic states. The presence of a Zeeman field $M_z$ offers spin splitting between the spin-up and spin-down states that leads to a spin-dependent DOS and transport properties. The Zeeman field can be provided either through either proximity coupling with a ferromagnetic substrate or an external magnetic field. The band structure gains spin-polarization as an effect, leading to spin-selective transmission channels and the possibility to realize spin filtering or magnetoresistive effects.

Fig. \ref{fig3} illustrates the variation of the energy resolved transmission probability $T(E)$ and the corresponding DOS with energy for ZGNR based device by varying out-of-plane magnetization strengths $M_z$, across two representative geometries: a narrower ribbon $(W,L) = (4 \mathrm{nm}, 12 \mathrm{nm})$ and a wider one $(8 \mathrm{nm}, 16 \mathrm{nm})$. For $M_z = 0$, the transmission is symmetric and shows pronounced peaks, indicating spin-degenerate edge localized states near the Fermi energy $E_F$, consistent with the presence of conducting edge states characteristic of pristine ZGNR, as already seen in Fig. \ref{fig2}. These quasi-flat bands contribute to a high DOS at $E_F$ and quantized steps in transmission, signifying the presence of perfectly conducting edge channels. In presence of a finite Zeeman-type interaction $M_z$ the spin degeneracy is lifted, resulting spin-splitting of the degenerate edge states. The corresponding DOS reflects this via an asymmetric but two distinct VHS centered about $E = 0$. The magnitude of this splitting grows significantly for $M_z = 1$, confirming the progressive spin polarization of edge modes.
The transmission spectra reveals a suppression of conductance near $E_F$ for finite $M_z$. This is due to the formation of a spin-dependent gap and the breaking of spin channel symmetry. With the increase in $M_z = 1$, the system tends toward a half-metallic phase, where only one spin species remains conductive within a finite energy window. This indicates a single spin channel that contributes to transmission and makes the system an efficient spin filter. We found that the narrower ribbons show more pronounced energy level quantization and enhanced exchange-induced splitting as seen from Figs. \ref{fig3}(a) and \ref{fig3}(c). This is due to stronger quantum confinement and an increased spatial overlap of edge-localized wavefunctions. In contrast, wider ribbons exhibit smoother DOS profiles and more gradual transmission transitions, as seen from Figs. \ref{fig3}(b) and \ref{fig3}(d). It reflects the reduced confinement and increased delocalization of edge states, thereby diminishing the impact of $M_z$. Moreover, the evolution of the DOS with increasing $M_z$ reveals a systematic broadening and eventual suppression of the peak at $E = 0$. This is originated from topologically non-trivial, sublattice-polarized edge states in the unpolarized case. The introduction of magnetization not only breaks time-reversal symmetry but also modifies the sublattice-resolved electronic structure. Thus, the degeneracy of the edge modes have been lifted. As a result, the spectral weight is redistributed to higher or lower energies depending on spin orientation. This redistribution is especially pronounced in the wider ribbon [Fig. \ref{fig3}(d)], where the higher density of available edge and bulk modes facilitates stronger mixing and mode hybridization. Our findings demonstarate the tunability of spin-resolved transport in ZGNR based device via Zeeman field and highlight the key role of edge state manipulation in designing graphene-based spintronic architectures. 

\begin{figure}[t]
\vspace{-5mm}
\centerline{ 
\includegraphics[scale=0.28]{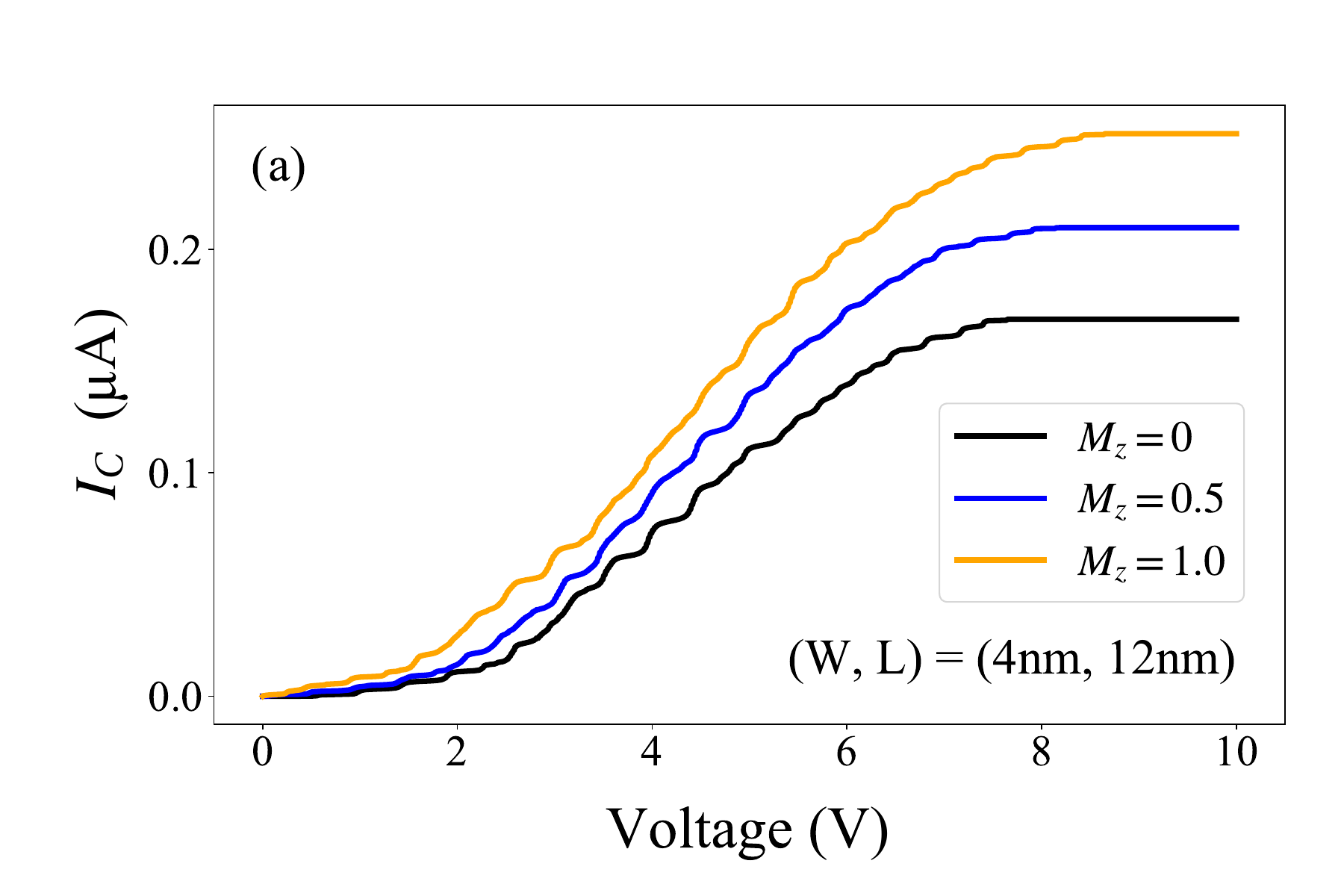}}
\vspace{-5mm}
\includegraphics[scale=0.28]{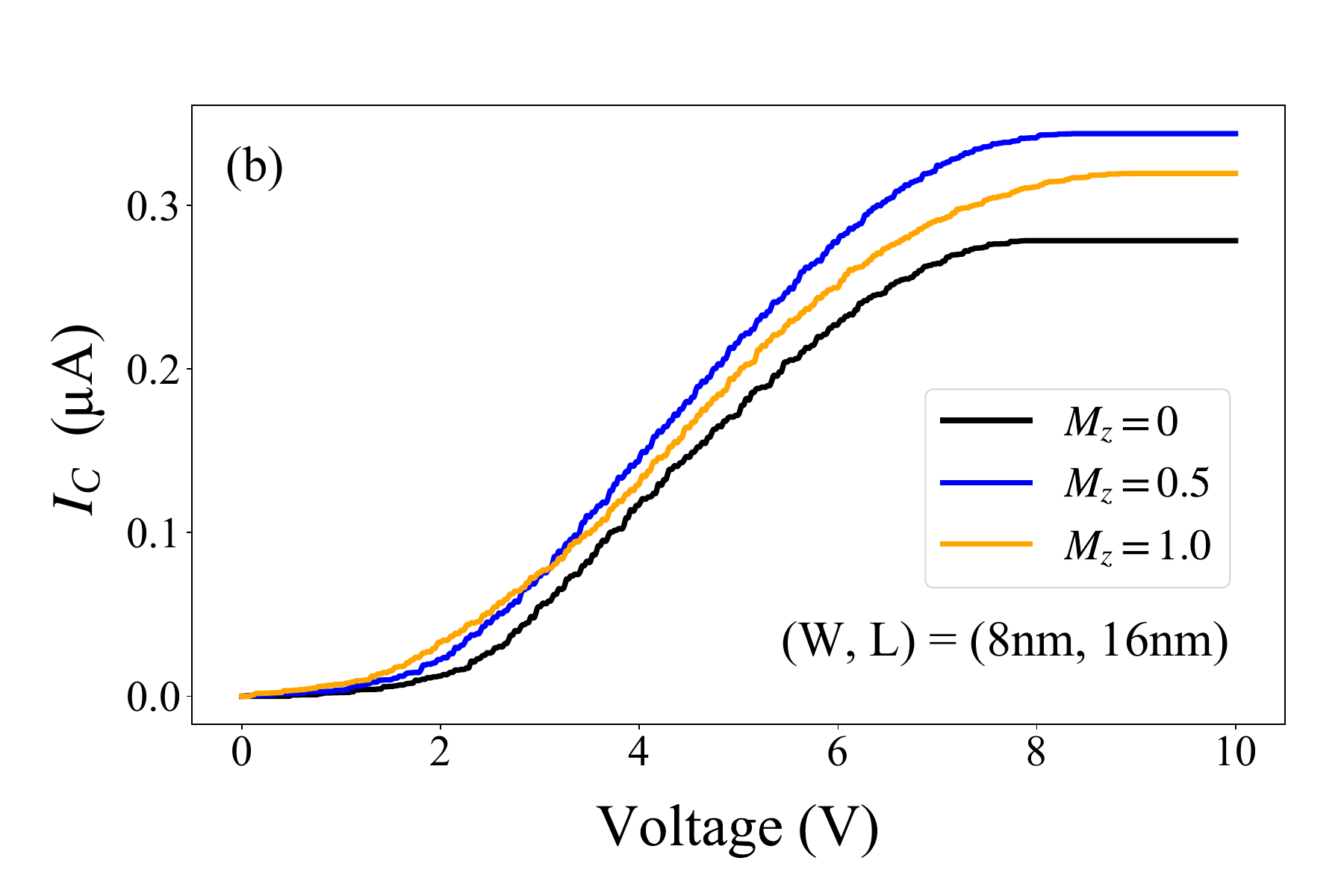}
\caption{I-V characteristics for different magnetization strength  considering  (a) $(W, L) = (4 \mathrm{nm}, 12 \mathrm{nm})$ and (b) $(W, L) = (8 \mathrm{nm}, 16 \mathrm{nm})$ of ZGNR. }
\label{fig4}
\end{figure}

Fig. \ref{fig4} shows the I-V characteristics for different magnetization strengths $M_z$ for two distinct device geometries: panel (a) corresponds to a narrower and shorter ribbon $(W,L) = (4\,\mathrm{nm}, 12\,\mathrm{nm})$  while panel (b) features a wider and longer device (b) $(W,L) = (8\,\mathrm{nm}, 16\,\mathrm{nm})$. We observe a clear transport gap under low bias conditions for both geometries. It indicates the quantum confinement and edge state in finite-length ZGNR. With an increase in magnetization, the current has been significantly increased indicating enhanced spin-resolved transmission.  This trend is a direct consequence of spin splitting induced by $M_z$ and activates additional conduction channels. The highest current amplitudes occur for $M_z = 1.0$, where ferromagnetic edge ordering facilitates spin-aligned transport through partially localized edge states, increasing the DOS near the Fermi level. In contrast, we observe intermediate currents in the $M_z = 0.5$ regime. This indicate the partial polarization and reduced edge alignment for narrow and shorter ribbons, as seen in Fig. \ref{fig4}(a). However, for longer and wider ribbons, the current for $M_z = 0.5$ exceeds that for $M_z = 1$, suggesting a non-trivial dependence of spin transport on both the device geometry and edge-state overlap.  In particular, the magnitude of current shows a gradual increase for $M_z \neq 0$  compared to that of $M_z = 0$. When comparing the two geometries, the wider ZGNR in Fig. \ref{fig4}(b) exhibits higher currents due to the availability of more number of transverse modes and reduced confinement effects. Furthermore, the smoother current onset in Fig. \ref{fig4}(b) suggests weaker edge localization, resulting in more delocalized, continuous transport paths. The nonlinearity in all I-V curves is the result of bias-induced shift in lead chemical potentials. This dynamically reshapes the transport window and selectively enables or suppresses conduction channels. These results collectively highlight the interplay between magnetization, edge state and quantum confinement in governing spin-dependent charge transport in ZGNR-based nanodevices.

\begin{figure*}[t]
\centerline{ 
\includegraphics[scale=0.28]{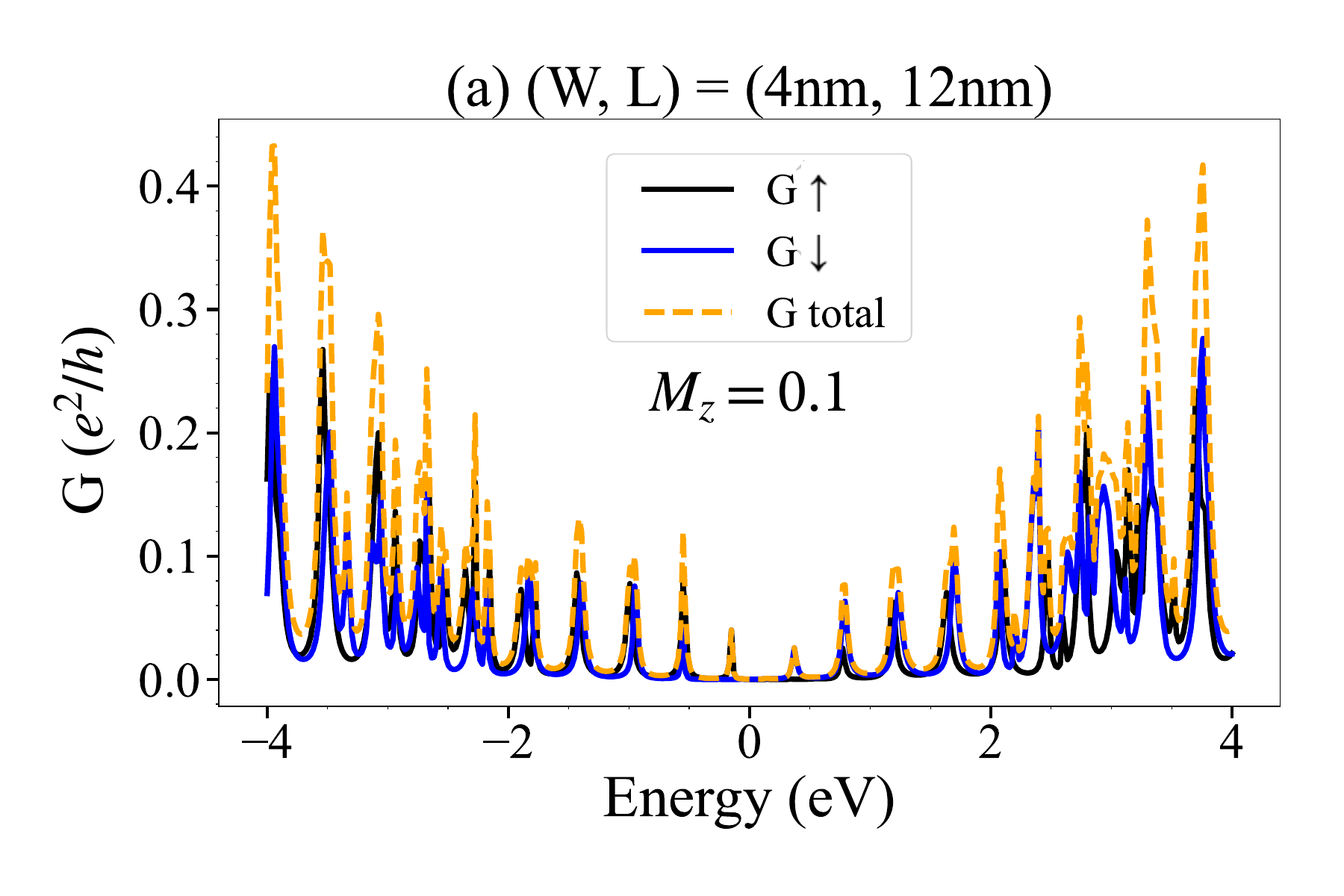}
\includegraphics[scale=0.28]{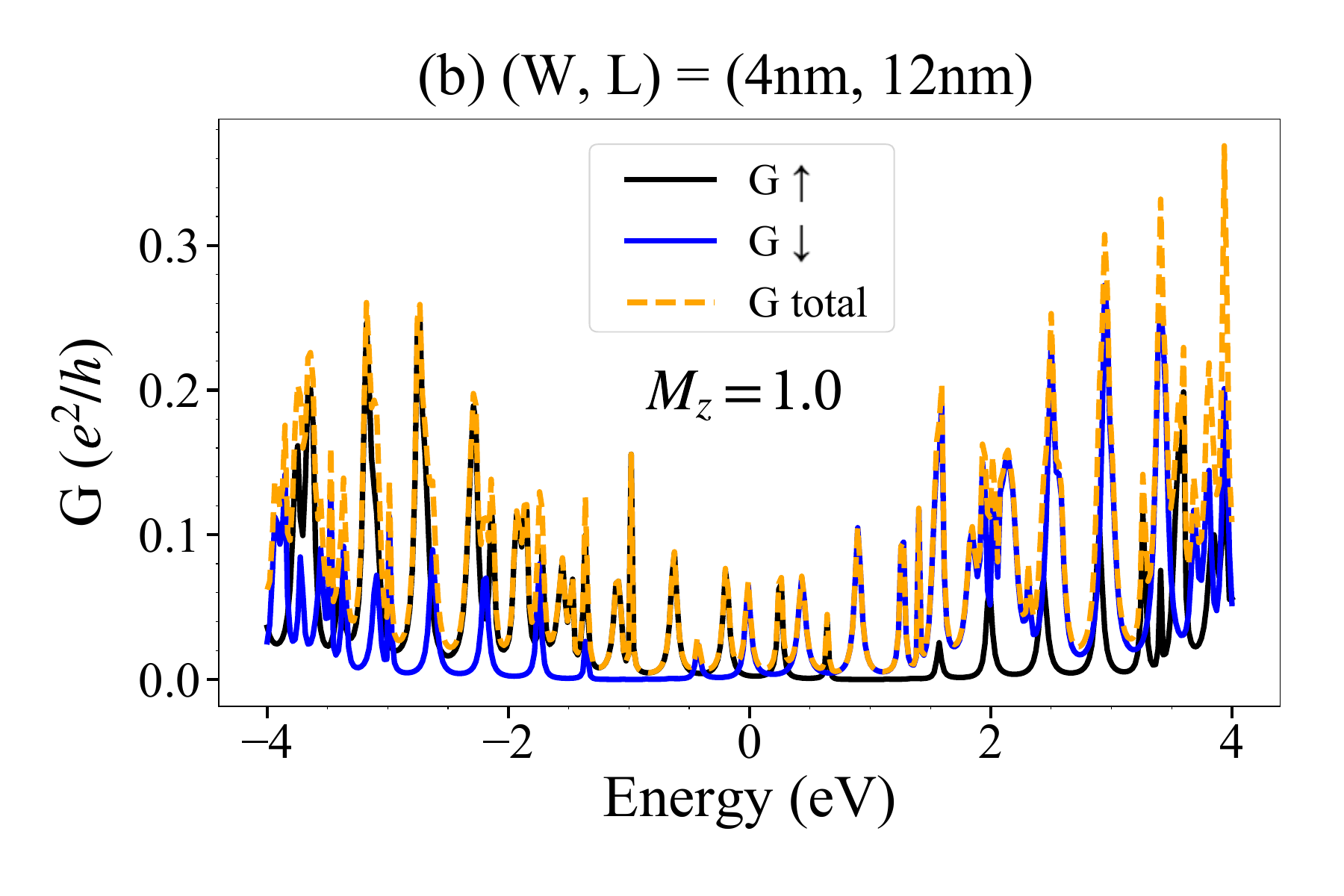}}
\caption{Variation of normalized conductance with energy    considering Zeeman strength (a) $M_z = 0.1$ and (b) $M_z = 1.0$. The plots are for ZGNR dimension $(W, L) = (4 \mathrm{nm}, 12 \mathrm{nm})$.}
\label{fig5}
\end{figure*}

\subsection{Spin resolved conductance}
Fig. \ref{fig5} shows the energy dependence of the spin resolved conductance $G^{\uparrow}$, $G^{\downarrow}$ and total conductance $G$ in a finite length ZGNR of dimensions $W = 4\,\mathrm{nm}$ and $L = 12\,\mathrm{nm}$, under two different Zeeman field strengths: $M_z = 0.1$ [Fig. \ref{fig5}(a)] and $M_z = 1.0$ [Fig. \ref{fig5}(b)]. These plots are computed using the Eq. (\ref{eq15}) and present the interplay between spin polarization, quantum confinement, and subband quantization in ZGNRs. In Fig. \ref{fig5}(a), for a relatively weak Zeeman field $M_z = 0.1$, the spin splitting is small but appreciable. The spin-up and spin-down conductance channels are slightly offset from one another in energy, particularly at higher bias values. This asymmetry results in the development of spin-resolved features, with $G^{\uparrow}$ and $G^{\downarrow}$ exhibiting non-overlapping resonant peaks. These peaks correspond to the quantized energy levels of the finite-length ribbon and governed by ribbon's geometry as well as the sublattice symmetry. The total conductance $G = G^{\uparrow} + G^{\downarrow}$ displays resonant structures that reflect the cumulative effect of both spin channels, yet still retains the fine structure due to weak spin splitting. However,  when the Zeeman strength is increased to $M_z = 1.0$, a strong spin asymmetry emerges, as seen in Fig. \ref{fig5}(b). Spin splitting becomes more promising when energetically decoupling the $G^{\uparrow}$ and $G^{\downarrow}$ channels. As a result, more pronounced separation of the conductance peaks, with distinct sets of resonances are observed for both the spin configuration. Near the Fermi level, one spin channel is significantly suppressed while the other remains conducting. It signifies spin-polarized transport and indicate its potential as a spin filter or spin valve. This arises mainly due to the broken particle-hole symmetry in the spectrum and is a direct consequence of both the Zeeman-induced spin splitting and the finite-size confinement effects. Moreover, the presence of multiple sharp peaks in both spin channels arises from Fabry-P\'erot-like interference within the finite ZGNR, where coherent multiple reflections at the ends of the nanoribbon give rise to discrete transmission resonances. These peaks broaden and shift differently for spin-up and spin-down carriers under stronger Zeeman fields. Overall, this figure demonstrates the transition from weak to strong Zeeman coupling highlights the controllable nature of spin splitting and the direct manifestation in the conductance spectra.

\begin{figure}[t]
\vspace{-5mm}
\centerline{ 
\includegraphics[scale=0.26]{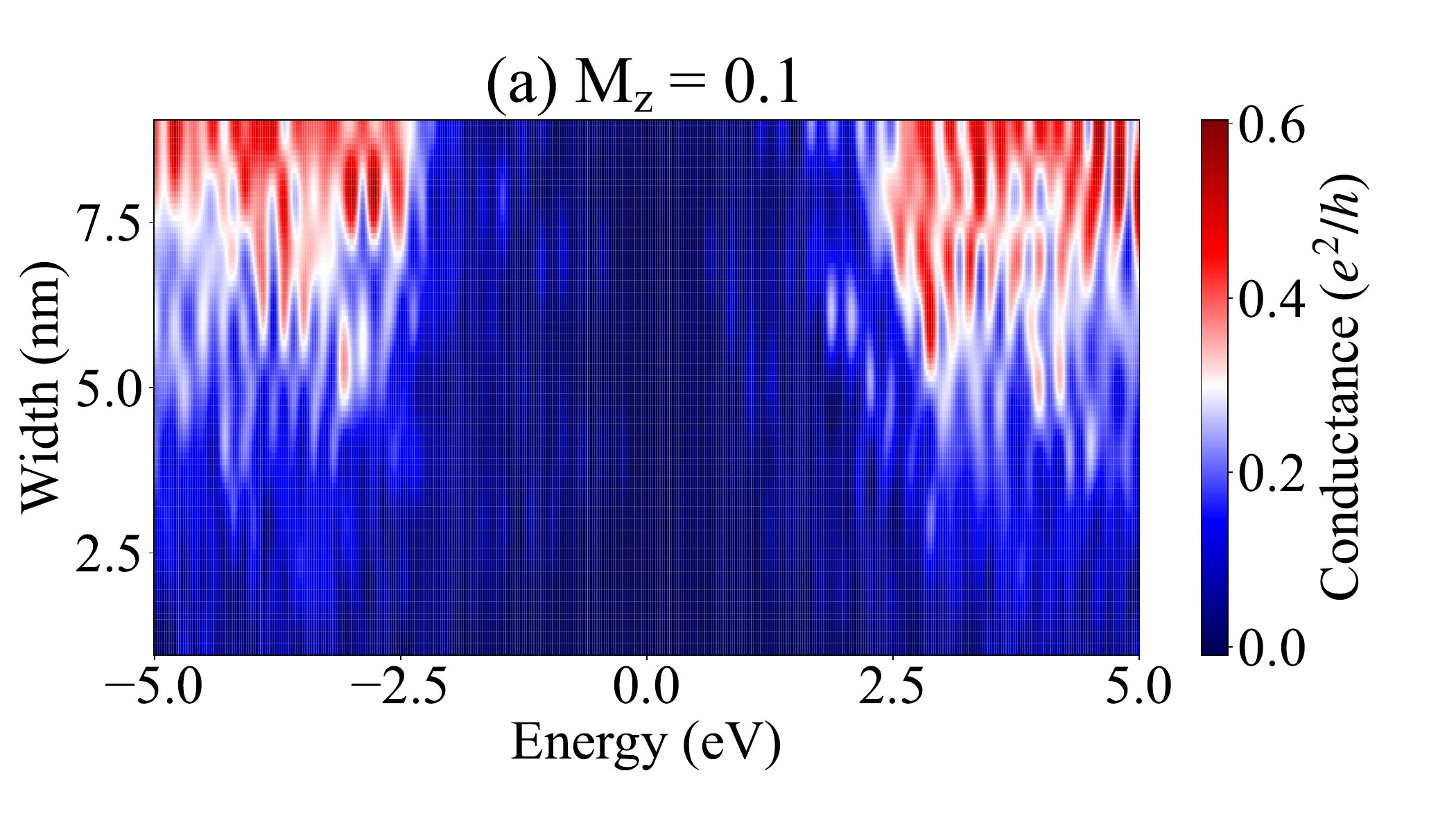}}
\vspace{-5mm}
\centerline{
\includegraphics[scale=0.26]{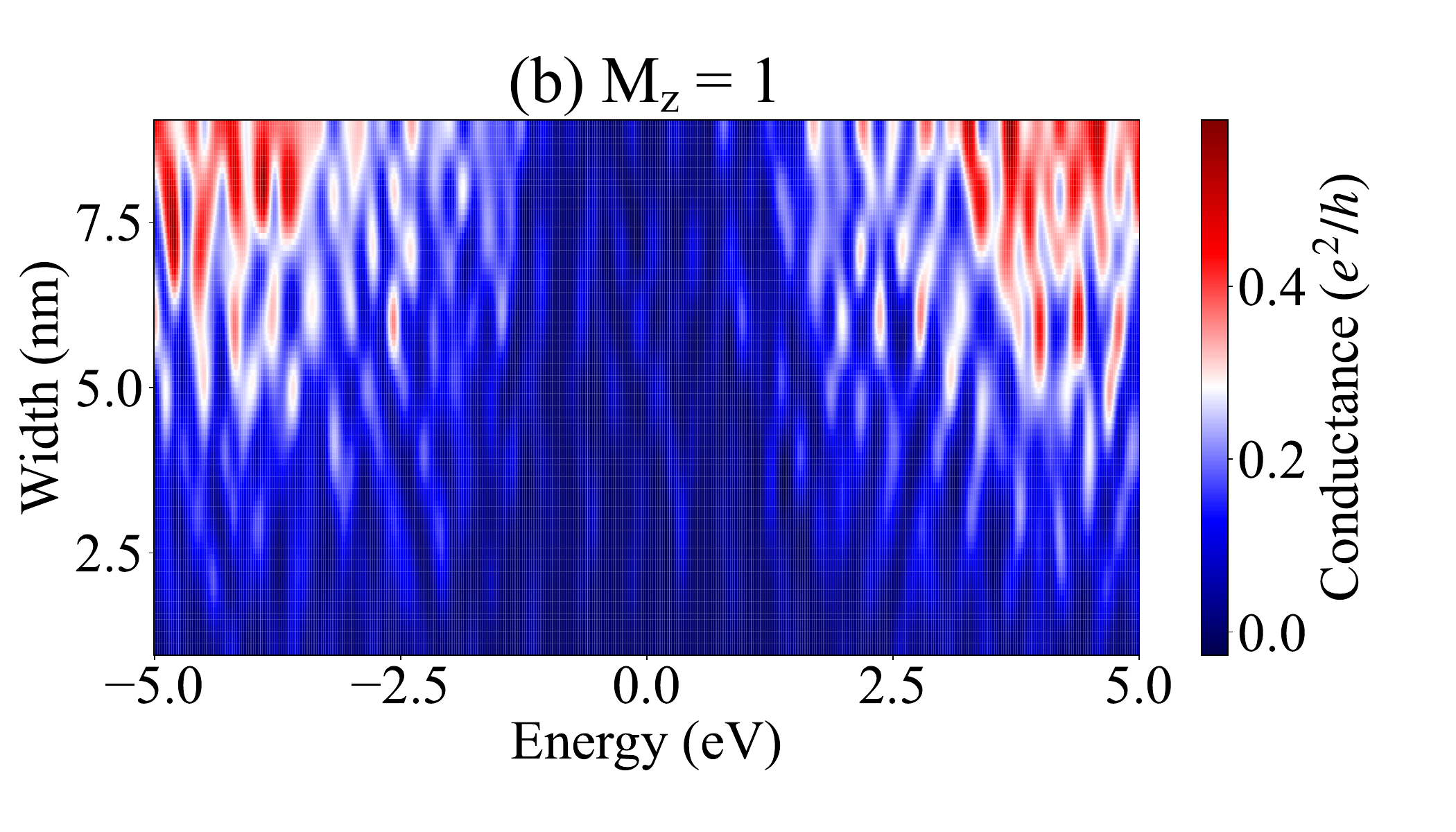}}
\caption{Density plot of variation of normalized conductance with energy and ZGNR width   considering Zeeman strength (a) $M_z = 0.1$ and (b) $M_z = 1.0$.}
\label{fig6}
\end{figure}

In Fig. \ref{fig6}, we present density plots of the normalized conductance as a function of energy $E$ and the width $W$ of the ZGNR channel for the Zeeman field strength: $M_z = 0.1$ [Fig. \ref{fig6} (a)] and $M_z = 1.0$ [Fig. \ref{fig6} (b)].  For low Zeeman strength $M_z = 0.1$, we observe enhanced conductance at energies near the band edges and for larger ZGNR widths ($W \gtrsim 6.5$~nm), indicating dominant transport through spin-degenerate edge states. The reason is because that the ZGNR edge modes are confined and produce distinct conducting channels, which are most noticeable as bright streaks in the conductance map and are caused by quantized transverse subbands. However, it is observed that near the charge neutrality point ($E = 0$), the conductance is significantly suppressed, particularly for narrower ribbons. This behavior is related to increased inter-edge coupling and finite-size-induced hybridization of opposite edge states, which open a transport gap due to the emergence of antiferromagnetic correlations or weak spin polarization.

\begin{figure}[t]
\vspace{-5mm}
\includegraphics[scale=0.27]{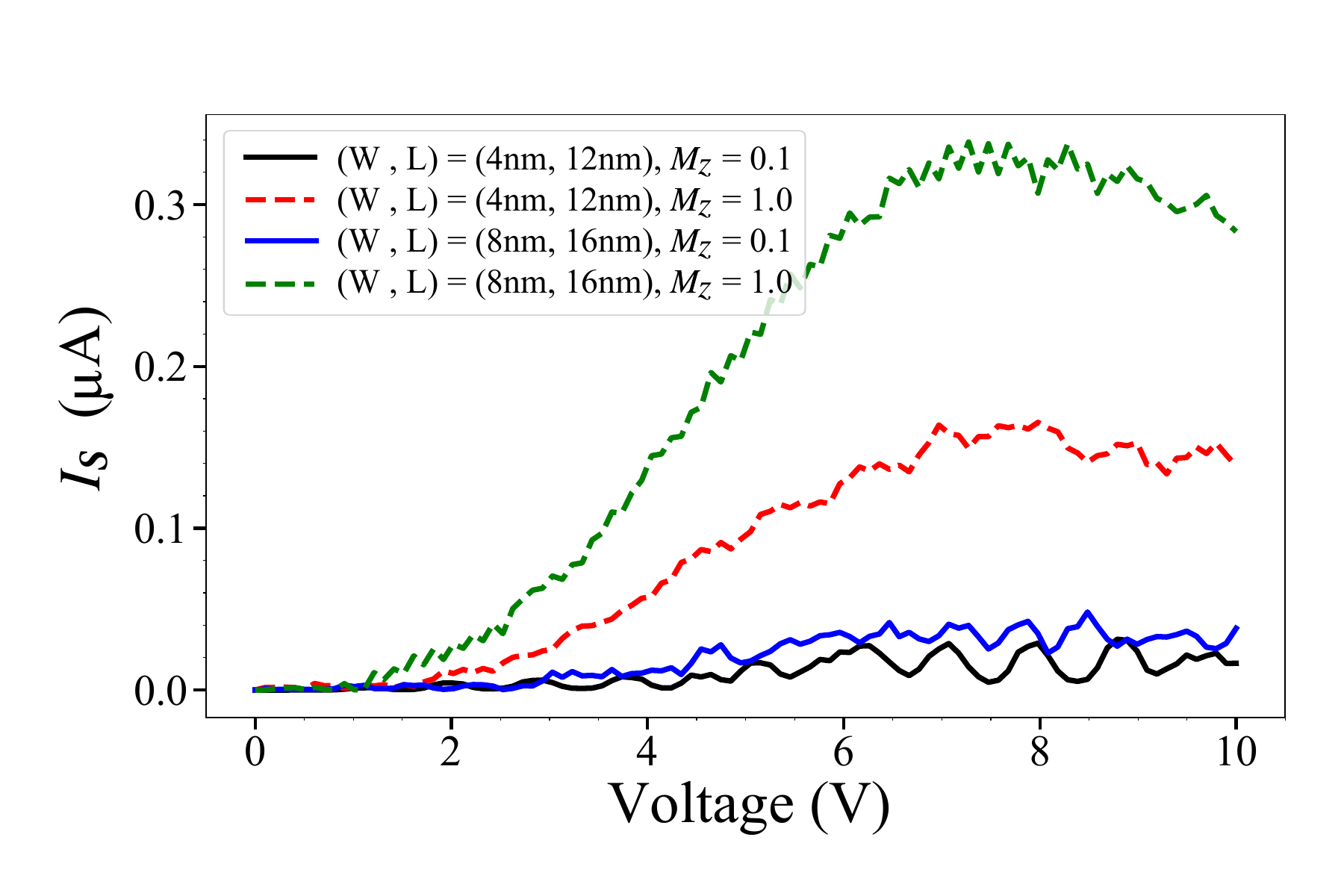}
\caption{Variation of  Spin current with biased voltage   considering weak ($M_z = 0.1$) and  strong ($M_z = 1.0$) Zeeman strength. The plots are for ZGNR dimension: $(W, L) = (4 \mathrm{nm}, 12 \mathrm{nm})$ and $(8 \mathrm{nm}, 16 \mathrm{nm})$.}
\label{fig7}
\end{figure}

As the Zeeman field increases to $M_z = 1.0$, the conductance spectra shows a qualitative transformation. In this condition, the spin degeneracy is lifted and the conductance shows highly asymmentric characteristics around $E = 0$ due to the opening of spin-dependent transport gaps and new channels. Moreover, in this condition narrow ribbons can also show significant conductance as seen in Fig. \ref{fig6} (b).  In addition, the prominent spectral weight shifts away from the Fermi level, indicating that only one spin channel contributes significantly to conduction within certain energy windows. This appears as asymmetric suppression and band fragmentation of the strong conductance bands, exhibiting selective filtering behavior of the spin-up or spin-down carriers. Moreover, the shift of the dominant conductance peaks to higher energies provide additional evidence for Zeeman-induced breaking of the particle-hole symmetry. Notably, the emergence of fine oscillatory features across the width is a signature of Fabry–P\'{e}rot-like resonances in the confined ZGNR region, modulated by spin-split subband structures. 

These results collectively demonstrate that the application of a Zeeman field not only breaks spin degeneracy but also offers controllable manipulation of spin-polarized edge channels. Thus, the conductance modulation with an energy-dependent width can be used to obtain the desired efficiency of spin filtering, providing a path toward fabricating scalable, gate-controllable and spin-selective transport devices using ZGNRs. In a three-terminal geometry, the effects come into play especially for spin-polarized current routing and logic functions, where the third terminal can be used for spin injection, extraction, or detection. The stability of these characteristics over a large range of energies and ribbon widths accentuates the practicality of achieving functional spintronic architectures with ZGNRs under modest magnetic fields, either by magnetic proximity coupling or by direct exchange with a ferromagnetic substrate. Our results thus identify the essential role of Zeeman-induced spin splitting in engineering the quantum conductance of three-terminal ZGNR-based devices.

\begin{figure}[t]
\vspace{-6mm}
\centerline{ 
\includegraphics[scale=0.28]{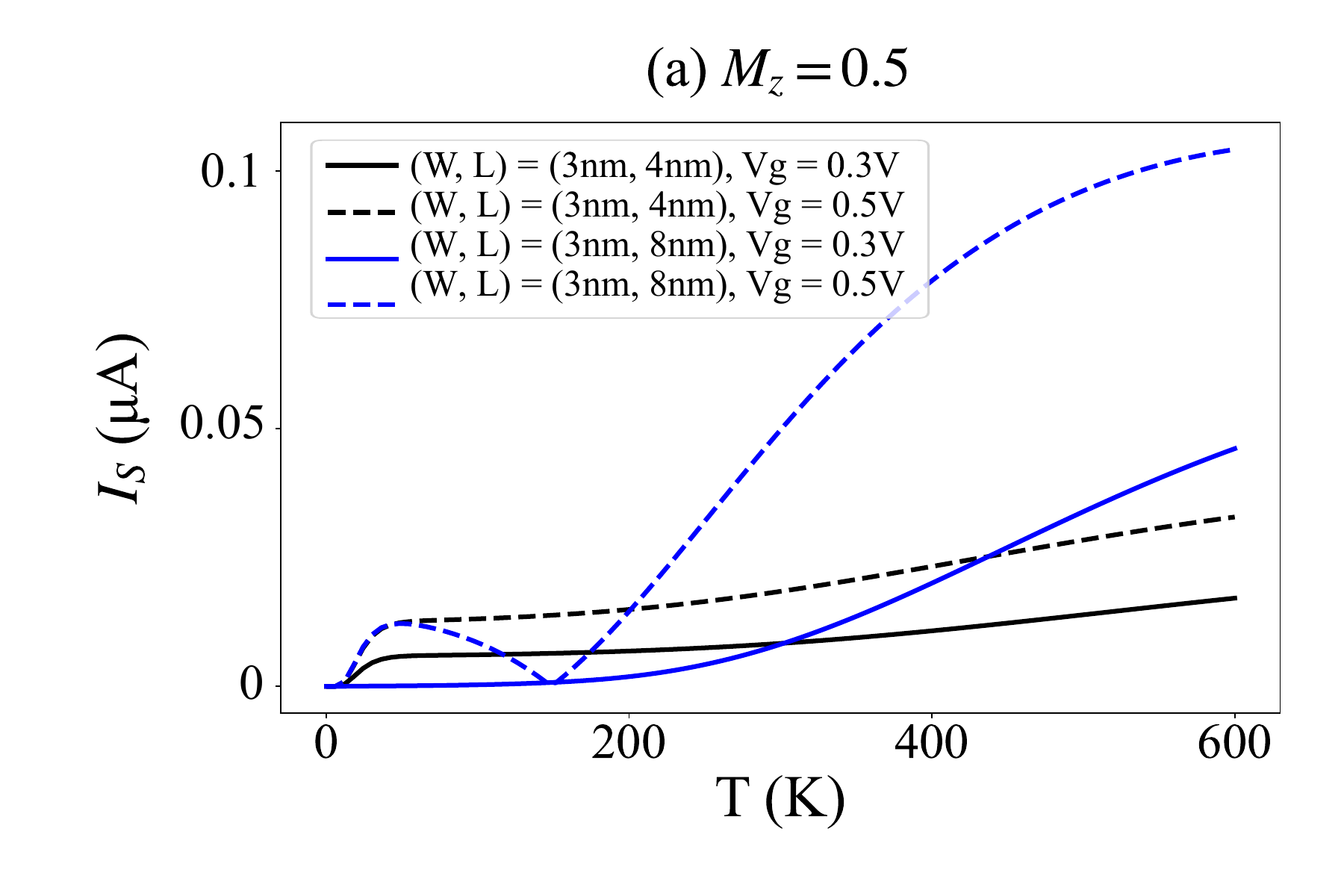}}
\vspace{-5mm}
\centerline{
\includegraphics[scale=0.28]{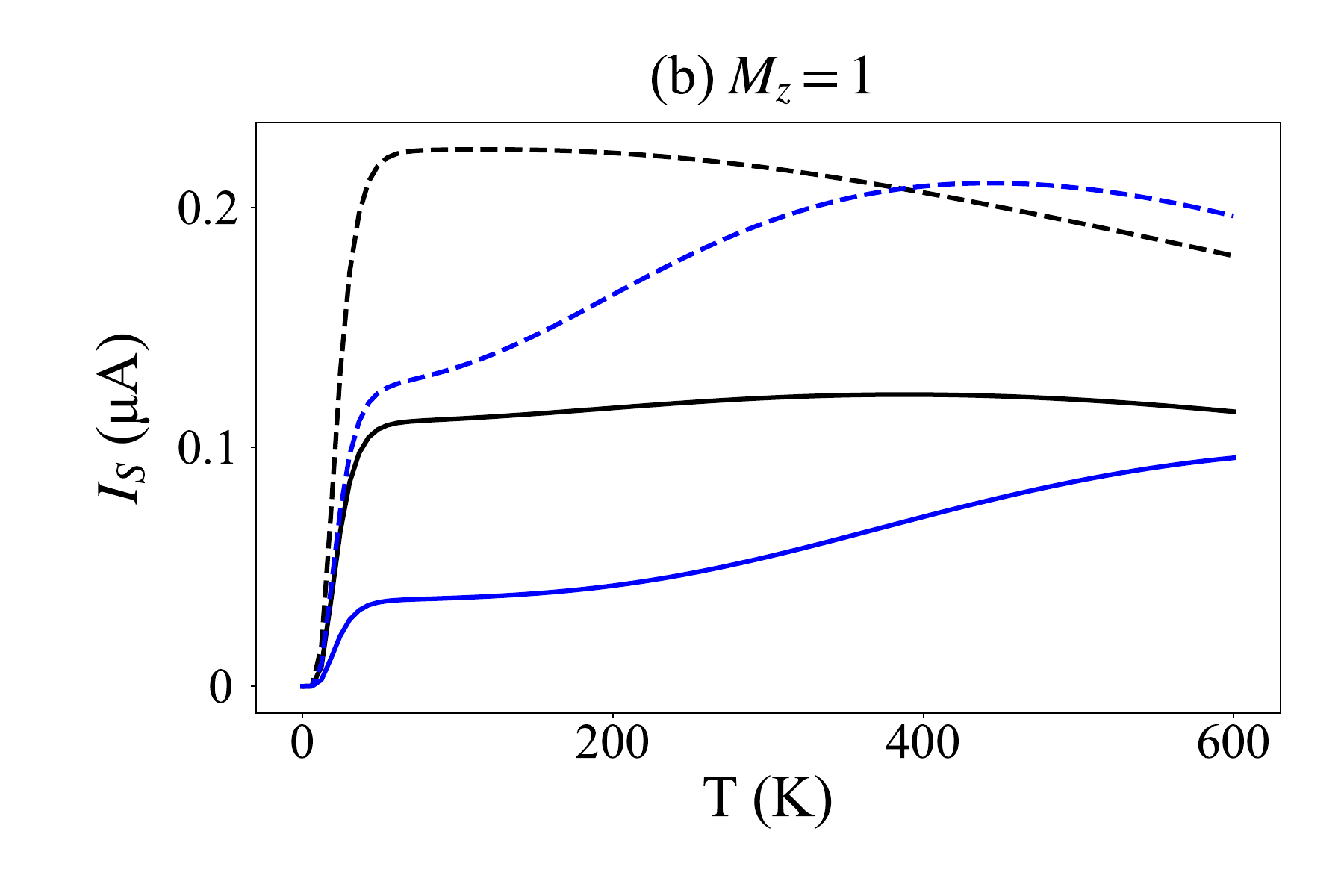}
}
\caption{Variation of Spin current with temperature  for (a) weak ($M_z = 0.1$) and  (b) strong ($M_z = 1.0$) Zeeman strength. The plots are for ZGNR dimension: $(W, L) = (3 \mathrm{nm}, 4 \mathrm{nm})$ (black line) and $(W, L) = (3 \mathrm{nm}, 8 \mathrm{nm})$ (blue line).}
\label{fig8}
\end{figure}

\begin{figure*}[t]
\centerline{ 
\includegraphics[scale=0.28]{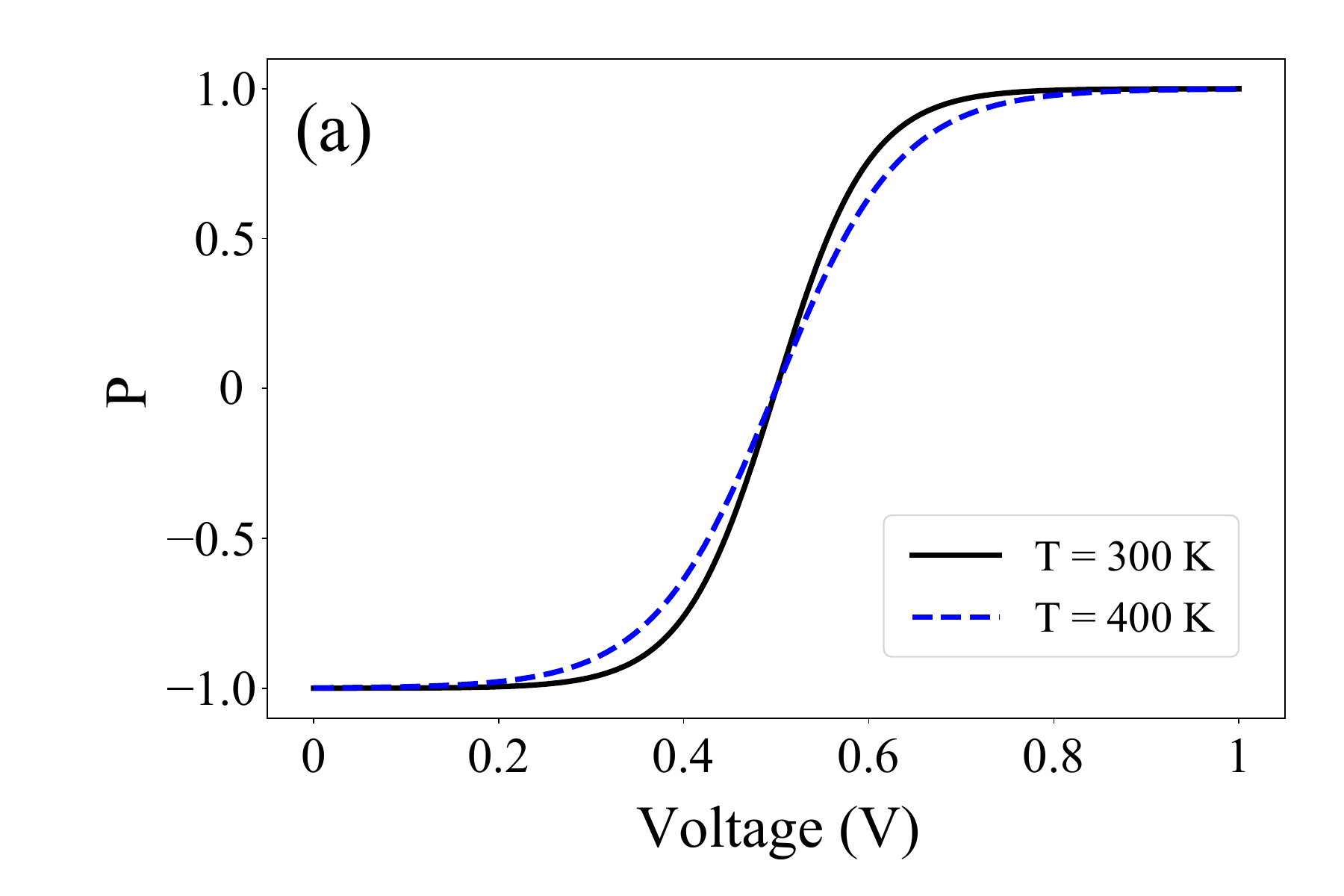}
\includegraphics[scale=0.28]{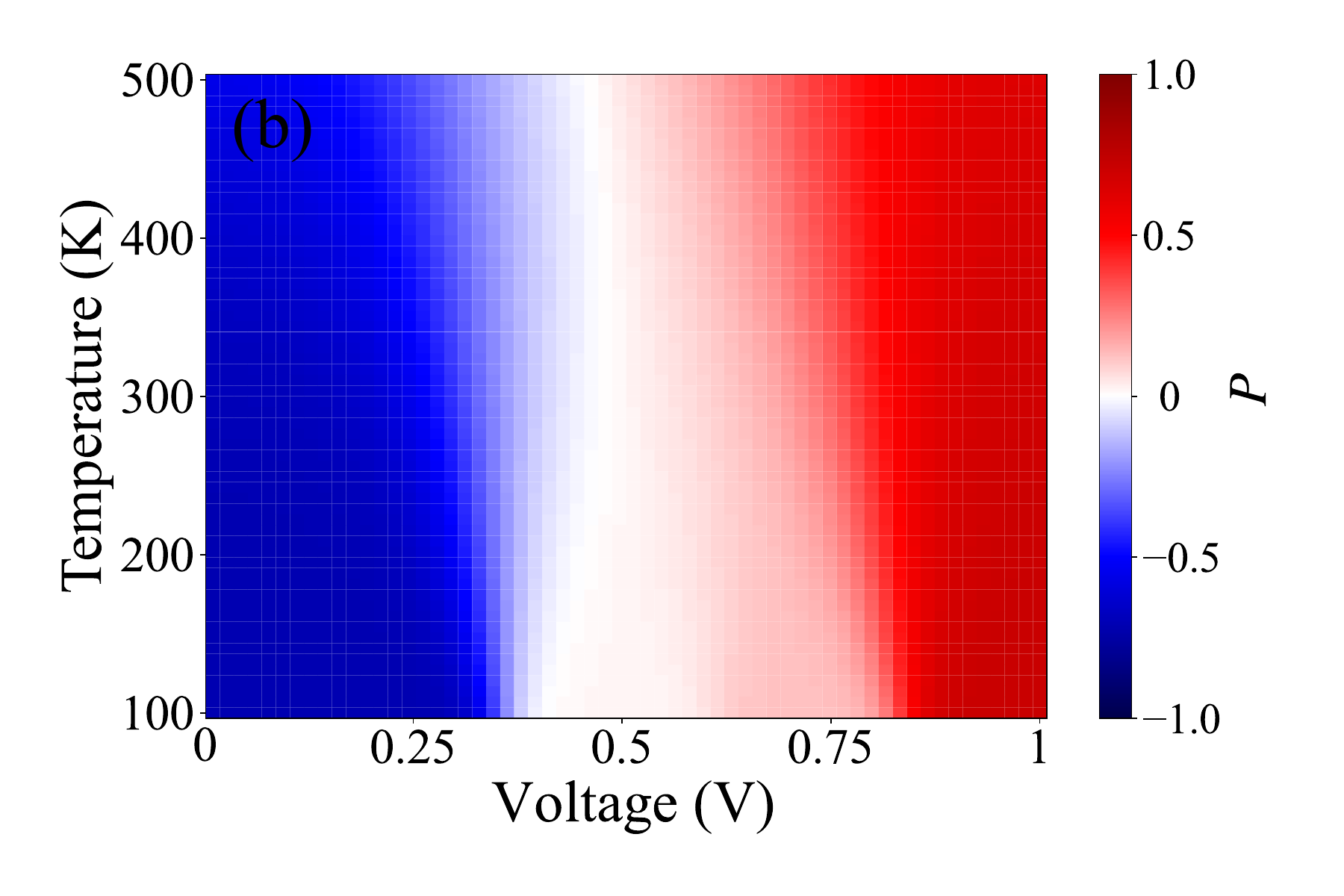}
}
\caption{Variation of Spin polarization with bias voltage for different temperatures. }
\label{fig9}
\end{figure*}
\subsection{Spin Current}
It is evident from Figs. \ref{fig5} and \ref{fig6} that transport through a three-terminal ZGNR is spin-dependent and can be tuned via an external Zeeman field. This motivates our investigation of the spin current, which refers to the net flow of spin angular momentum and can be quantified as the difference between spin-up and spin-down currents. In Fig. \ref{fig7}, we present the evolution of spin current $I_S$ as a function of applied bias voltage using Eq. (\ref{eq18}) for two ZGNR geometries: $(W, L) = (4\,\mathrm{nm}, 12\,\mathrm{nm})$ and $(8\,\mathrm{nm}, 16\,\mathrm{nm})$, under weak ($M_z = 0.1$) and strong ($M_z = 1.0$) Zeeman fields. We observe that the spin current shows a gradual increase with bias voltage for $M_z = 1$ for both ribbon geometries indicating a clear dependence on the Zeeman field strength, demonstrating the tunability of spin-polarized transport through magnetic exchange interaction.  For $M_z = 0.1$, spin splitting is minimal, and contributions from spin-up and spin-down nearly cancel out, resulting in a weak net spin current. Notably, for the narrower ribbon ($W = 4\,\mathrm{nm}$), the spin current rises linearly at low bias and saturates at higher voltages, indicating dominant transport through a single spin channel due to a finite-size-induced spin gap and suppressed inter-subband mixing. In contrast, the wider ribbon ($W = 8\,\mathrm{nm}$) exhibits low spin current across all voltages even at $M_z = 1.0$ due to enhanced subband mixing and partial cancellation between spin-polarized channels arising from edge-state hybridization. The presence of oscillatory characteristics in the spin current, particularly at low $M_z$, can be attributed to Fabry-P\'{e}rot-type quantum interference from multiple reflections within the finite-length nanoribbon. The overall behavior confirms that both the ribbon width and the Zeeman energy  act as effective control parameters for engineering spin-polarized current in ZGNR-based three-terminal devices. These results indicate that narrow ZGNRs with strong Zeeman fields can function as efficient spin injectors or spin filters in ballistic spintronic circuits, where gate voltage and magnetic field provide versatile external control. Thus, the voltage-tunable spin current response makes these systems promising for developing logic architectures and low-power spintronic interconnects.

Fig. \ref{fig8}, illustrates the variation of $I_S$ with temperature for different ZGNR geometries under (a) weak ($M_z = 0.5$) and (b) strong ($M_z = 1.0$) Zeeman fields. Furthermore, we consider the effect of the gate voltage by considering $V_g = 0.3V$ and $0.5V$. For weaker Zeeman field $M_z = 0.5$, the spin current increases gradually with temperature across all device geometries as seen in Fig. \ref{fig8}(a). This behavior is attributed to thermal activation and enables higher-energy spin-polarized states to contribute to transport.
In particular, longer ribbons ($L = 8\,\mathrm{nm}$) yield a larger spin current than shorter ones due to enhanced spin splitting over extended lengths, thereby facilitating spin-selective conduction. Furthermore, increasing $V_g$ enhances $I_S$ by shifting the chemical potential into regions with a greater spin-resolved DOS asymmetry. Moreover, we observe that $I_S$ saturates in the higher $T$ region for a shorter ribbon. In contrast, Fig. \ref{fig8}(b) shows that for a stronger Zeeman field $M_z = 1.0$, the spin current rises sharply at low temperatures and then saturates. This saturation originates from the substantial spin splitting that has already been present due to the strong Zeeman field. In particular, thermal effects only marginally enhance the current beyond a certain threshold temperature. Once thermal broadening exceeds the spin gap, both spin channels contribute and the spin current asymmetry stabilizes. Importantly, the saturation value of $I_S$ is significantly higher for longer ribbons and higher $V_g$, consistent with the trends observed in Fig. \ref{fig8}(a). These results also demonstrate that temperature and gate voltage act as effective tuning parameters for spin-polarized transport in ZGNRs. Also, longer ribbons exhibit enhanced spin filtering effects owing to their greater sensitivity to spin-dependent potentials.

\begin{figure*}[t]
\centerline{
\includegraphics[scale=0.26]{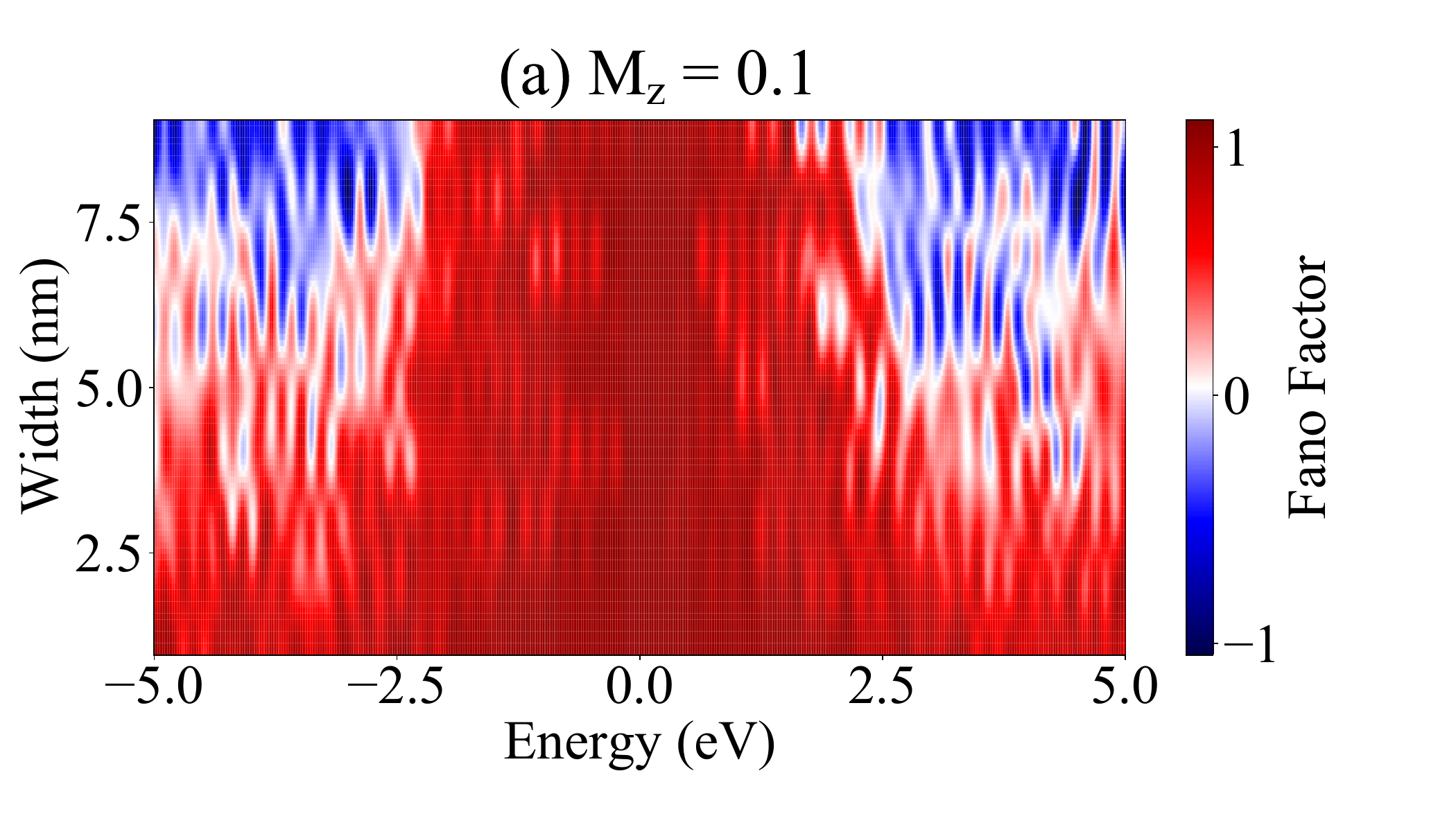}
\hspace{-5mm}
\includegraphics[scale=0.26]{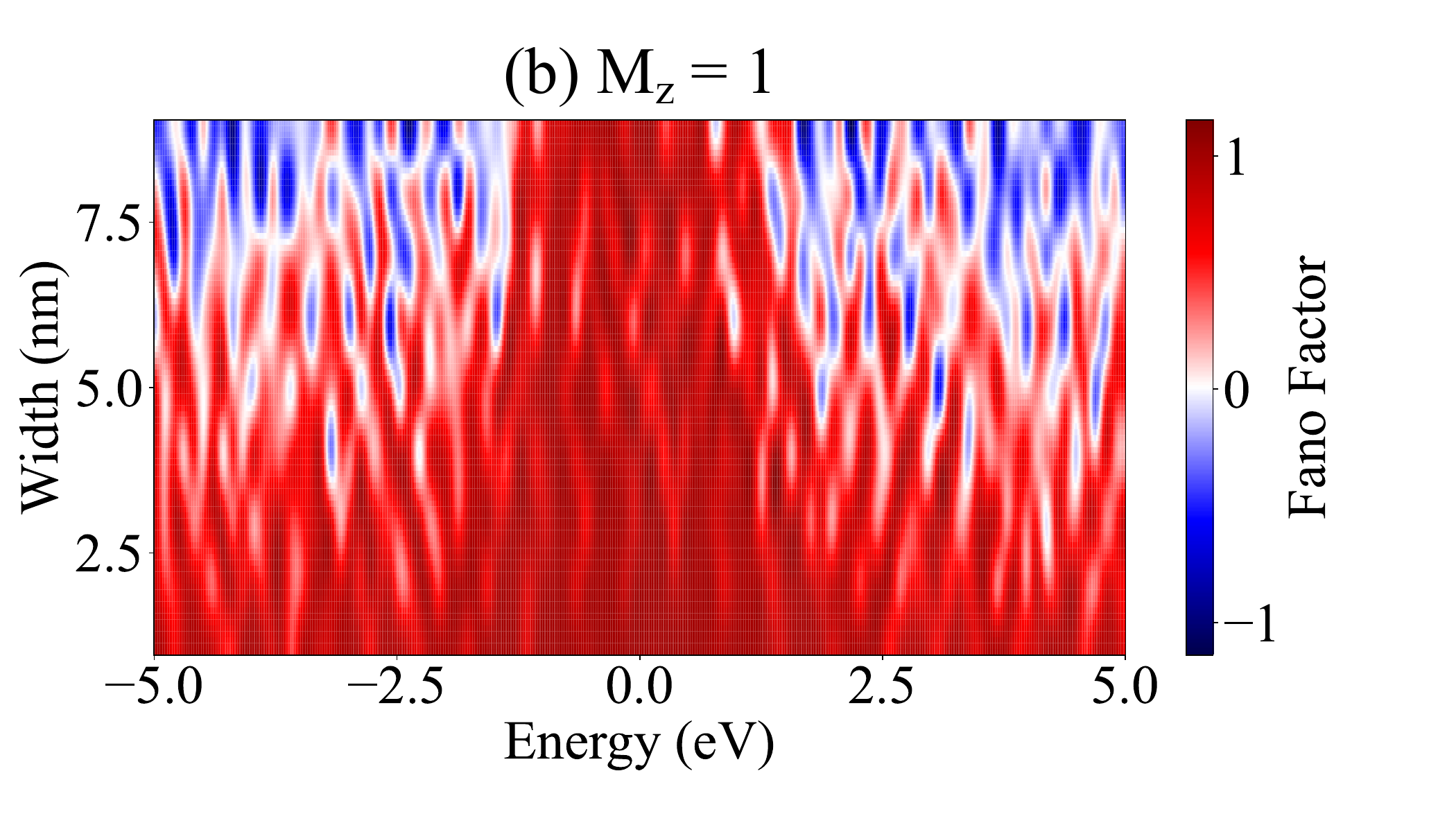}}
\caption{Density plot of  Fano factor with energy and ZGNR width.    considering Zeeman strength (a) $M_z = 0.1$ and (b) $M_z = 1.0$. }
\label{fig10}
\end{figure*}

\subsection{Spin polarization}
Spin polarization is a degree of measure of the imbalance between spin-up and spin-down currents. It can be defined as \cite{datta2015}
\begin{equation}
\mathcal{P} = \frac{I^{\uparrow}(V,T) - I^{\downarrow}(V,T)}{I^{\uparrow}(V,T) + I^{\downarrow}(V,T)},
\label{eq19}
\end{equation}
where, $\mathcal{P} = +(-) 1$ correspond to fully polarized spin-up (spin-down) configuration respectively, while $\mathcal{P} = 0$ indicates equal contributions from spin-up and spin-down currents. Thus, $\mathcal{P}$ quantifies the net spin imbalance in the transport current and serves as a key indicator of spin-selective conduction. In addition, understanding the dependence of $\mathcal{P}$ on bias voltage and temperature is significant in the realization of spin-filtering effect.

The bias-voltage dependence highlighting the effect of temperature of the spin polarization  
for ZGNR device is illustrated in Fig. \ref{fig9}. 
We show the variation of $\mathcal{P}$ with voltage for two representative temperatures, 
$T = 300~\mathrm{K}$ (solid black line) and $T = 400~\mathrm{K}$ (dashed blue line) in Fig. \ref{fig9}(a), 
while Fig. \ref{fig9}(b) shows a colormap of $\mathcal{P}$ as a function of both bias voltage and temperature.
We observe that $\mathcal{P}$ transit from the spin-down configuration ($\mathcal{P} \rightarrow  -1$) to the spin-up configuration ($\mathcal{P} \rightarrow 1$) with the increase in the bias voltage. The transition is found to be sharper for $T = 300 K$ than $T = 400K$ as seen in Fig. \ref{fig9}(a). This behavior originates from the energy-dependent spin-resolved transmission through the ZGNR, which is influenced by edge magnetism and the applied Zeeman field. The bias window symmetrically encompasses both spin species at $ V = 0.5V$, resulting in nearly equal spin-up and spin-down currents resulting $\mathcal{P} \approx 0$. As the bias voltage increased, the asymmetry in the transmission spectrum between spin-up and spin-down channels becomes prominent which result in spin-polarized transport through ZGNR channel. The temperature dependence shown in Fig. \ref{fig9}(b) reveals thermal broadening effects.  It is observed that at higher temperatures, the Fermi-Dirac distribution smears the occupation probabilities, reducing the sharpness of the transition in $\mathcal{P}$ and slightly changing the threshold voltage required for full polarization. Thus, the magnitude of polarization decreases for a given bias configuration, as observed in the comparison between $300\,K$ and $400\,K$ in Fig. \ref{fig9}(a). Moreover, the polarization remains robust over a wide temperature range despite thermal effects. Thus, it suggest the potential applicability of ZGNR-based spintronic devices at room temperature and beyond.

\subsection{Fano factor}
To understand the nature of transport, quantum correlations and fluctuations of charge carriers in a transport system, it is necessary to investigate the Fano factor, which defined as the ratio of shot noise to Poissonian noise and can be obtained by the relation \cite{acharjee2024}
\begin{equation}
F_{\alpha\beta} = \frac{\int_{-\infty}^{\infty} \left( -\frac{\partial f(E)}{\partial E} \right) T_{\alpha\beta}(E)\left[1 - T_{\alpha\beta}(E)\right] \, dE}{\int_{-\infty}^{\infty} \left( -\frac{\partial f(E)}{\partial E} \right) T_{\alpha\beta}(E) \, dE}
\label{eq20}
\end{equation}

Fig.  \ref{fig10} shows the density plots of the Fano factor as a function of energy and ZGNR width for Zeeman field strengths: (a) $M_z = 0.1$ and (b) $M_z = 1.0$. 
In Fig. \ref{fig10}(a), in the weak Zeeman regime $M_z = 0.1$, the Fano factor shows significant fluctuations mainly around the band edges ($E  \approx \pm 2.5$ eV) and beyond. These oscillations signifies the resonant backscattering and increased mode mixing, occurring as a result of interplay between edge-localized spin-polarized states and the spin-conserving scattering potentials within the system. The quantum size effects responsible for the strong modulations in the Fano factor with respect to ribbon width are evident from the dominance of discrete transverse subbands and spin interference effects in the transport dynamics. Specifically, the strongly enhanced sensitivity of the Fano factor to the ribbon width in this regime implies the suppression of current due to interference, in agreement with the appearance of Fano-like resonances generated by localized edge states that interfere with the extended modes. 

Upon increasing the Zeeman field to $M_z = 1.0$, as illustrated in Fig.\ref{fig10}(b), the entire pattern of the Fano factor becomes more symmetrized and structured. The areas with large positive and negative values of the Fano factor increase and become more uniform in the energy range. This is a result of increased spin splitting of the subbands caused by the strong Zeeman coupling, which creates stronger spin-filtering effects and spin-selective transmission. The modulation of the Fano factor thus corresponds to spin-polarized edge state channels becoming energetically well-separated, suppressing spin mixing and resulting in partially suppressed noise in certain energy and width ranges. In particular, the Fano factor at the Fermi level becomes less fluctuating with increasing width, meaning a more stabilized transport channel dominated by spin-polarized states with suppressed backscattering.

\section{Summary and conclusions}
In this work, we have explored spin-resolved ballistic transport in a three-terminal Zigzag Graphene Nanoribbon (ZGNR) based device under varying geometries and Zeeman field strengths. We employ non-equilibrium Green function approach and Landauer-B\"{u}ttiker formalism to study the transmission spectrum, density of states, I–V characteristics, spin-resolved conductance, spin current, spin polarization and the Fano factor for three-terminal ZGNR device. We found that transport is governed by edge-localized states and subband quantization, with resonant features strongly influenced by ribbon width and length in absence of magnetization.  The application of an out-of-plane Zeeman field lifts spin degeneracy and thus enables spin-selective transport, leading to the emergence of half-metallic characteristics and efficient spin filtering, particularly in narrower ribbons. The conductance spectra and I–V characteristics reveal strong spin-dependent features, including Fabry–P\'{e}rot-like interference, bias-induced mode activation and geometry-dependent current asymmetry.  The spin current is found to be tunable by both magnetic field and gate voltage with enhanced response in narrow ribbons under strong magnetization. Temperature effects induce a saturation behavior in spin current and broaden the spin polarization transition, yet maintain robustness over a broad thermal range, making our ZGNR based device highly suitable for room-temperature applications. Finally, the dependence of Fano factor on the energy and width exhibits quantum interference signatures and spin filtering effects, reinforcing the role of spin coherence and mode hybridization. Altogether, our results demonstrate the high tunability and scalability of three terminal ZGNR-based devices for spintronic applications.  It offers insights for designing logic circuits, spin injectors and magnetoresistive elements with controllable spin polarization using modest magnetic fields.

\end{document}